\documentclass[letterpaper,12pt]{article} % DO NOT CHANGE THIS
\usepackage[margin=1in]{geometry}
\usepackage[hyphens]{url}  % DO NOT CHANGE THIS
\usepackage{graphicx} % DO NOT CHANGE THIS
\usepackage{caption} % DO NOT CHANGE THIS AND DO NOT ADD ANY OPTIONS TO IT
\usepackage{algorithm}
\usepackage{algorithmic}
\usepackage{amsmath}
\usepackage{amsfonts}
\usepackage{tikz}
\usepackage{standalone}
\usepackage{comment}
\usepackage{tcolorbox}
\usepackage{subcaption}
\usepackage{amsthm}
\usepackage{microtype}
\usepackage{placeins}

\usetikzlibrary{shapes.geometric, arrows.meta, positioning, calc, backgrounds}

\newtheorem*{remark}{Remark}
\theoremstyle{definition}

\newcommand{\kate}[1]{{\color{red}{KL: #1}}}

\usepackage{booktabs}

\title{Contracting for LLM Delegation: Moral Hazard in Technology and Effort Choice}

\author{
    Nanda Kishore Sreenivas, Kate Larson \\
    University of Waterloo \\
    \{nksreenivas, kate.larson\}@uwaterloo.ca
}

\date{}

\begin{document}
\maketitle

\begin{abstract}
We extend the standard Principal-Agent framework to scenarios where the Agent selects from a suite of technologies, each characterized by a distinct cost-capability profile. This framework is increasingly critical in the era of Large Language Models (LLMs), where Agents choose both a model and an associated effort level (e.g., token budget). We model the relationship between output quality and effort as a concave, saturating function, which depends on the Agent's hidden two-dimensional action choice balancing technology selection and effort allocation. We derive the optimal linear contract for the Principal, demonstrating that the Agent’s best response is characterized by a threshold reward share that triggers technology switching. Finally, we calibrate our model using open-weight LLM pairings across the MATH and MMLUPro benchmarks. We show that both Principal and Agent, when employing bandit algorithms to navigate this environment, converge to strategies that closely align with our theoretical equilibrium. These results suggest that simple linear contracts can effectively incentivize complex, technology-aware delegation in agentic workflows.
\end{abstract}


\section{Introduction}

The rapid capability improvements in large language models (LLMs) in recent years have fundamentally changed how tasks are executed: from reactive, single-prompt chat interfaces toward delegation to autonomous agents. Principals, whether individual users or firms, now routinely hand off complex, open-ended tasks to specialized AI Agents. This is evident in the rise of both generalized and domain-specific tools, e.g., legal firms delegate due diligence to systems like Harvey~\cite{harvey2026}, software teams assign bug fixes to autonomous coding agents like Devin~\cite{devin2026}, etc. In these interactions, the Agent operates as a black box; the Principal provides the objective and receives the final output, completely blind to the internal reasoning process and choices. 
%healthcare providers utilize Hippocratic AI for patient outreach

As this ecosystem matures, delegation will increasingly occur not just from humans to AI, but from AI to AI across open `agentic markets'. Orchestrator algorithms will dynamically delegate specialized sub-tasks to agents to minimize computation costs or leverage domain expertise \cite{hadfield_economy_2025, tomasev_intelligent_2026}. In these agent-oriented markets, the two transacting parties possess distinct economic identities and potentially misaligned utilities~\cite{rauba2026multiagentsystemstreatedprincipalagent}.

This economic reality exposes an issue with the current standard of ``pay-per-token" or pay-for-compute API pricing. When an autonomous Agent is billed based on its computational effort, it creates a severe moral hazard. As Bu and Ma note~\cite{bu_position_2026}, 
%\kate{I think this is OK. The observation is not crazy}
%(\nanda{this is just a workign paper}), 
token-based pricing under-incentivizes hidden effort and misaligns the Agent's objectives with those of the Principal. An Agent paid per token is incentivized to maximize verbosity, unnecessarily ``overthink" simple problems, or covertly utilize cheaper, lower-capability underlying models to save on its own costs~\cite{velasco2026auditing, saig_adaptive_2026}. 
%We are not the first to claim this; people have already pointed it out. Not sure, that is coming through in this para. See quotes: 

%Exact quote from the Ma paper: "token-based pricing under-incentivizes hidden effort, such as verification, tool use, and model selection, thereby creating moral hazard"

%Exact quote from Saig: "As inference is performed internally and a fixed price is agreed upon in advance, firms can strategically increase their profit margin by generating text using a cheaper, lower-quality model"

Consequently, delegating a task to an autonomous Agent increasingly means delegating a choice of \emph{which} tool to use, not just how hard to work. This is a moral hazard problem, but not one the classical Principal-Agent literature typically addresses. 
%Standard models let an Agent choose how much effort to exert under a single, fixed technology. 
We formalize this with a linear contract (parametrized by $\alpha \in [0,1]$) between a Principal and an Agent who selects a model $m$ and a token budget $x$, where output quality follows a saturating, diminishing-returns production function in $x$ which is a natural fit for LLM inference.
%, and also used in the anytime algorithm literature. 
Under this model, we characterize the Agent's best response as a threshold in the linear contract at which the Agent's optimal model choice switches between models, and we also derive the Principal's optimal linear contract.  We calibrate the production function on six pairings of open-weight models spanning three model families, across two task domains (MATH and MMLUPro), and show that a Principal and Agent using simple online learning algorithms converge to contracts and best responses close to our theoretical predictions.

\begin{comment}
\kate{You don't need the text below. The preceeding paragraph does a nice summary, and some of these points in too detailed for the introduction.}
Our contributions are: (1) a Principal-Agent model of joint technology and effort choice under moral hazard (2) the Agent's best response, characterized by a switching threshold between technologies, and the Principal's optimal linear contract; (3) relevant extensions in the LLM domain such as considering burn-in tokens, fixed payments in addition to linear contracts, costly verification; and (4) an empirical validation calibrating the model to real LLM production curves and showing that simple bandit algorithms recover behaviour consistent with theory.
\end{comment}

\begin{comment}
    
\begin{enumerate}
    \item[\textbf{RQ1}] For a given linear contract, what is the two-dimensional best response for the Agent?
    \item[\textbf{RQ2}] What is the optimal linear contract for the Principal?
    \item[\textbf{RQ3}] Can a \emph{learning} Principal and an Agent reach the expected equilibrium using actual open-weight LLMs on real-world tasks?
\end{enumerate}
\end{comment}

\section{Related Work}
\label{sec:rel_work}
%\kate{This can obviously be cut back signifcantly -- not by dropping references but just by cutting back on the text.}

Our work sits at the intersection of several areas: classical and algorithmic contract theory, anytime algorithms, inference-time compute allocation, and LLM mechanism design.

%The classical \textbf{principal-agent model} \cite{holmstrom_moral_1979} studies how a principal can incentivize an agent whose action or effort is hidden. Algorithmic contract theory \cite{duetting_algorithmic_2024} extends this to discrete, combinatorial action spaces and multiple outcomes, but treats the technology matrix (the probability of each action leading to an outcome) as exogenous and fixed for the design problem. A related strand of algorithmic contract theory studies online/bandit learning of contracts when the Principal does not know the Agent's action space, costs, or outcome distribution in advance~\cite{bacchiocchi_learning_2024, bacchiocchi_regret_2025, zhu_sample_2023}. This line of work focuses on the algorithmic and statistical question of \emph{how fast} a principal can learn a good contract, typically over discrete action and outcome spaces. Existing work, including the multitask model of \cite{holmstrom_multitask_1991}, treats the agent's production technology as fixed or decomposes a single technology into multiple effort or task dimensions; none lets the agent choose \emph{which} technology governs the effort-to-outcome mapping.

The classical \textbf{principal-agent model} \cite{holmstrom_moral_1979} studies how a Principal can incentivize an Agent whose action or effort is hidden. Algorithmic contract theory \cite{duetting_algorithmic_2025} extends this to discrete, combinatorial action spaces and multiple outcomes, but treats the technology matrix (the action-outcome probabilities) as exogenous and fixed. A related thread studies online learning of contracts focusing on questions of learnability and regret, typically over discrete action and outcome spaces~\cite{zhu_sample_2023, bacchiocchi_learning_2024, bacchiocchi_regret_2025}. Existing work, including the multitask model of Holmstrom and Milgrom~\cite{holmstrom_multitask_1991}, decomposes a single technology into multiple effort or task dimensions; none lets the Agent choose \emph{which} technology governs the effort-to-outcome mapping.

The \textbf{anytime algorithm} literature studies systems that can be interrupted at any point, returning an output whose quality improves with computation time \cite{zilberstein_using_1996, horvitz_87_uai}. This relationship is formalized via \emph{performance profiles}, concave and monotonically increasing functions mapping allocated compute to expected output quality. We borrow this object to model the production function of an LLM. This literature typically treats resource allocation as a single-agent deliberation~\cite{horvitz1990ideal,boddy_solving_1989}.
%:the party choosing the compute budget and the party consuming the output are the same, so there is no hidden action and nothing to contract over. 
%We instead place performance profiles in a contracting environment. 
Since our model uses common performance profiles from this literature, our results apply here too.
%, not only LLMs.

%Recent LLM engineering research investigates how LLMs can dynamically scale their \textbf{inference-time compute} based on task difficulty. AnytimeReasoner \cite{qi_optimizing_2025} trains a single model via RL (Budget Relative Policy Optimization) to produce a usable answer at any token budget, sampling budgets during training rather than optimizing only a fixed final length; related approaches similarly train models to regulate their own reasoning length \cite{wen_budgetthinker_2025, alomrani_reasoning_2025}, and document diminishing and eventually saturating returns to additional tokens. We rely on this literature's empirical findings to motivate the shape of our continuous production curves. There are also recent frameworks for model routing with the objective of minimizing inference costs: BEST-Route uses query difficulty to pick a model and the number of responses to aggregate~\cite{ding_best-route_2025}, UniRoute proposes feature-based dynamic routing~\cite{jitkrittum_universal_2025}, ARES for effort selection in multi-step tasks~\cite{yang_ares_2026}.  
%But the objective in this area is not economic: the RL objective aligns the model's own token allocation with the training signal, so there is no second party and no incentive-compatibility question. 

Recent LLM research investigates how LLMs can dynamically scale their \textbf{inference-time compute} based on task difficulty. AnytimeReasoner \cite{qi_optimizing_2025} trains a single model via RL to produce a usable answer at any token budget; related approaches similarly train models to regulate their own reasoning length \cite{wen_budgetthinker_2025, alomrani_reasoning_2025}, and document diminishing and eventually saturating returns to additional tokens. We rely on these findings to motivate the shape of our production curves. There are also recent frameworks for model routing with the objective of minimizing inference costs from a single-agent perspective~\cite{ding_best-route_2025, jitkrittum_universal_2025, yang_ares_2026}.

Our work also relates to the emerging application of \textbf{mechanism design to LLMs}. Much of this literature focuses on the seller's side, analyzing how an LLM provider screens buyers who have different task requirements \cite{bergemann_economics_2025, bergemann_menu_2026}. Dutting \emph{et al.}~\cite{dutting_mechanism_2024} explore token-level auctions to influence the output of LLMs for applications like ad generation. Contemporary work explores contracts over LLM generation, focusing on how and when the Principal should verify the Agent's work when verification is costly~\cite{saig_adaptive_2026}. Closest in spirit to our application is the work of Saig \emph{et al.}~\cite{saig_incentivizing_2024}, who motivate `pay-for-performance' contracts and design threshold and monotone contracts for an Agent choosing among a discrete menu of LLMs, robust to unknown Agent costs, over a discrete space of quality levels. We build on this premise of contracting over AI generation but alter the fundamental mechanics. The Agent is a task/domain specialist using publicly available models and, therefore, costs are known in our theoretical model. We also introduce an explicit continuous, unobservable token budget nested within the discrete model choice.

\section{Model}
\label{sec:model}

\begin{figure*}[t]
    \centering
    \begin{tikzpicture}[
    % Core block styling
    block/.style={
        rectangle, 
        draw=black!80, 
        fill=white,
        thick, 
        text width=3.2cm, 
        align=center, 
        minimum height=1.5cm,
        rounded corners=2pt,
        font=\small
    },
    % Arrow styling
    arrow/.style={
        -{Stealth[scale=1.2]}, 
        thick, 
        draw=black!70
    },
    % Information leakage / observability dashed lines
    obs_line/.style={
        draw=black!50, 
        dashed, 
        thick
    }
]

    % --- STAGE 1: Principal ---
    \node[block] (principal) {
        \textbf{1. Contract Stage} \\
        \vspace{0.1cm}
        Principal ($P$) offers linear contract $\alpha \in [0,1]$
    };

    % --- STAGE 2: Agent's Hidden Decisions ---
    \node[block, right=1.5cm of principal] (agent_model) {
        \textbf{2a. Model Choice} \\
        \vspace{0.1cm}
        Agent ($A$) selects \\ $m \in \{L, H\}$
    };
    
    \node[block, below=0.6cm of agent_model] (agent_effort) {
        \textbf{2b. Effort Allocation} \\
        \vspace{0.1cm}
        Agent ($A$) allocates token budget $x \geq 0$
    };

    % --- STAGE 3: Production ---
    \node[block, right=1.5cm of agent_model] (production) {
        \textbf{3. Execution} \\
        \vspace{0.1cm}
        LLM $m$ runs inference, yielding quality: \\ $q = q_m(x)$
    };

    % --- STAGE 4: Payoffs ---
    \node[block, right=1.5cm of production] (payoffs) {
        \textbf{4. Payoff Realization} \\
        \vspace{0.1cm}
        $U = \alpha q - c_m x$ \\
        $V = (1-\alpha)q$
    };

    % --- BACKGROUND BOX FOR AGENT'S INTERNAL STATE ---
    \begin{scope}[on background layer]
        \draw[draw=black!40, fill=gray!5, dashed, rounded corners=4pt] 
            ($(agent_model.north west)+(-0.2,0.4)$) rectangle ($(agent_effort.south east)+(0.2,-0.2)$);
        \node[text=black!70, font=\small\bfseries, above=0.5cm of agent_model] {Agent's Action Space (Hidden)};
    \end{scope}

    % --- CONNECTING ARROWS ---
    \draw[arrow] (principal.east) -- (agent_model.west);
    \draw[arrow] (principal.east) -- ($(agent_model.west)+(-0.6,0)$) |- (agent_effort.west);
    
    \draw[arrow] (agent_model.east) -- (production.west);
    \draw[arrow] (agent_effort.east) -| ($(production.west)+(-0.6,0)$) -- (production.west);
    
    \draw[arrow] (production.east) -- (payoffs.west);

    \draw[arrow, dotted] (agent_model.south) -- (agent_effort.north);
\end{tikzpicture}
    \caption{Principal-Agent Model with hidden model choice \emph{and} effort.}
    \label{fig:flowchart}
\end{figure*}
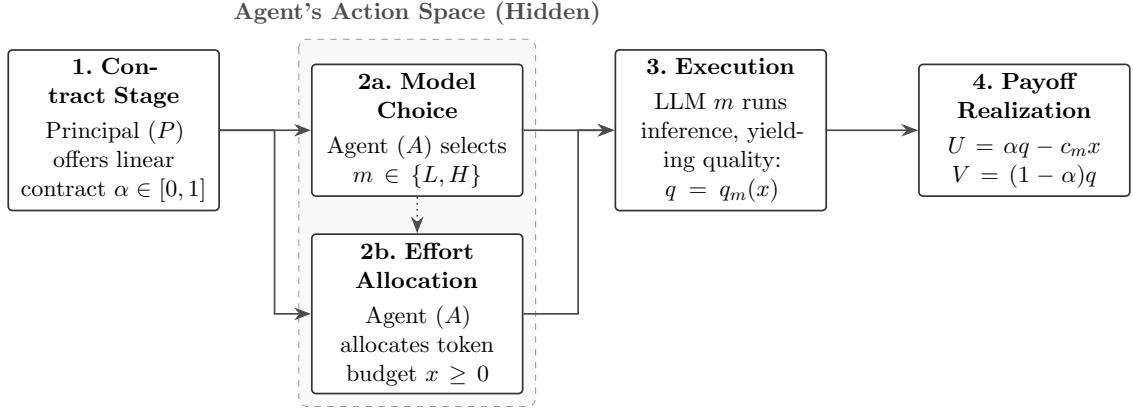

A \textbf{Principal} $P$ delegates a task to an \textbf{Agent} $A$. The Agent is an \emph{anytime reasoner}: it selects a language model $m$ from a finite set of models and allocates token budget $x \geq 0$ to inference. Output quality $q$ is a  function of $m$ and $x$, and only the quality is observable by $P$; not the Agent's model choice or token budget. We assume that the Agent specializes in a specific task domain, and therefore, all contracted tasks are reasonably similar. The overall setup is shown in Figure~\ref{fig:flowchart}.

The Principal offers a \emph{linear contract}\footnote{We explore fixed payments in Appendix C and show that, under limited liability, it is equivalent to this reward sharing model.} with reward share $\alpha \in [0,1]$. The Agent receives $\alpha \cdot q$ and the Principal retains $(1-\alpha)\cdot q$. The contract is over output $q$ only; model choice and token count are not directly observable, and therefore, not contractible. 

Let $\mathcal{M}$ be the set of available models, and models are arranged in increasing order of capability in the task domain. We generalize to multiple models in Appendix H. For clarity, we present the two-model scenario with $\mathcal{M} = \{L, H\}$, where $L$ represents the cheaper model with lower capability and $H$ denotes a more capable and more expensive model. For example, $L$ could be a smaller model that is optimized for edge applications, while $H$ could be an expensive, slower frontier model. Alternatively, $L$ could be a simple, instruction-tuned chat model while $H$ could have other capabilities such as reasoning, tool calls, etc. Each model $m \in \{L, H\}$ is characterized by a set of performance and cost parameters. 
 
We assume linear cost per token, and the output $q$ is measured in terms of accuracy (as a percent, $q \in [0,100]$)\footnote{Measuring quality of models is outside the scope of this paper.}. 
%\kate{In a footnote, it might make sense to include a sentence that says something that the details of measuring quality is outside the scope of the paper. You don't want a reviewer asking details about this.}

The Agent's two decisions are:
\begin{itemize}
  \item \textbf{Model choice:} $m \in \{L, H\}$.
  \item \textbf{Effort:} token count $x \geq 0$.
\end{itemize}

The utilities of the Agent and Principal are given by:
\[
  \bar{U} = \alpha \cdot v \cdot q_m(x) - \bar{c}_m x, \quad 
  \bar{V} = (1-\alpha)\cdot v \cdot q_m(x).
\]
where $m$ is the model choice and $\bar{c}_m$ is the true cost per token. We assume that the monetary return to the Principal scales linearly with accuracy by a factor $v$. We normalize utilities by dividing by the Principal's valuation parameter $v$ throughout\footnote{This assumes that the Principal's task valuation $v$ is public, which may not always be true. The information asymmetry over the Principal's `type' forces the Agent to `screen' the Principal. 
%through a menu of contracts (with different prices), 
We leave this for future work.} to get:
\begin{align}
    U(\alpha,m,x) &= \alpha \cdot q_m(x) - c_m x, \notag \\ 
    V(\alpha,m,x) &= (1-\alpha) \cdot q_m(x). \label{eq:utilities}
\end{align}

where $c_m$ denotes the normalized cost per token for model $m$, \emph{i.e.}, $c_m = \bar{c}_m/v$. We will use these normalized utilities for the Agent and Principal in the rest of this paper.

\begin{comment}
\begin{table}[h]
\centering
\begin{tabular}{ll}
\toprule
\textbf{Symbol} & \textbf{Meaning} \\
\midrule
$\alpha$              & Contract share in $[0,1]$ \\
$x$                   & Token budget allocated by the Agent \\
$q_m(x)$              & Accuracy of model $m$ given budget $x$ \\
$M_m, k_m$            & Capability ceiling and sat. rate of $m$ \\
$c_m$                 & Cost per token, model $m$ \\
$b_m$                 & Burn-in tokens, model $m$ \\
$\tau_m$              & Activation thres.: min. $\alpha$ for $m$ to be viable \\
$\tau_m^P$            & Burn-in-adjusted activation threshold \\
$\theta$              & Switching threshold \\
$\theta_0$            & Switching threshold under burn-in \\
$\alpha_m^*$          & Principal's unconstrained optimal $\alpha$ under $m$ \\
$\alpha_m^\dagger$     & Principal's constrained optimal $\alpha$ under $m$ \\
$U_m(\alpha)$         & Agent's utility under model $m$ \\
$V_m(\alpha)$         & Principal's payoff under model $m$ \\
%$S_m(\alpha)$         & Total surplus under model $m$ \\
\bottomrule
\end{tabular}
\caption{Notation summary [For our reference; TMP].}
\label{tab:notation}
\end{table}
\end{comment}

\subsection{Optimal Effort}

To further analyse the Agent's optimal model choice and effort level, we model the output function $q_m(x)$ as a saturation function, where $M_m$ is the capability ceiling (in terms of accuracy on the task) and $k_m$ is the saturation rate of model $m$.

\begin{align}
  q_m(x) = M_m\!\left(1 - e^{-k_m x}\right) \label{eq:saturation_func}
\end{align}
%\kate{You need to immediately say what $M_m$ and $k-m$ are after the equation.}
\begin{comment}
Note that the function is concave, increasing, and has an asymptotic upper bound $M_m$ which mirrors the capability ceiling of an LLM at a specific task domain. The saturation shape is consistent with empirical studies of LLM test-time compute: several papers on ``LLM overthinking" report that accuracy stagnates beyond a point and the marginal returns diminish at higher budgets at rates that vary across models and tasks~\cite{wen_parathinker_2025, zhou_when_2026}. The accuracy curves of RL-trained anytime reasoners that are explicitly built to produce a usable answer at any budget also exhibit a similar qualitative shape~\cite{qi_optimizing_2025}. Beyond LLMs, the anytime-algorithms literature has described performance profiles with diminishing marginal returns~\cite{zilberstein_using_1996}. Early anytime planner work ~\cite{boddy_solving_1989} modeled expected performance with a saturating exponential $Q(t) = 1-e^{-\lambda t}$.
\end{comment}

Note that the function is increasing, concave, and has an asymptotic upper bound $M_m$, and this shape is consistent with empirical studies of LLM inference. Several papers report that accuracy stagnates and the marginal returns diminish at higher token budgets~\cite{wen_parathinker_2025, zhou_when_2026, qi_optimizing_2025}. Beyond LLMs, the anytime algorithm literature has also used similar curves to model performance profiles~\cite{zilberstein_using_1996, boddy_solving_1989}. 

%\kate{You can cut back this last paragraph a bit, by just including the citations but without going into the details, if you need space}

%"For example, Boddy and Dean (1989) used the function $Q(t) = 1 - e^{-\lambda t}$ to model the expected performance of their anytime planner."  --- Zilberstein_96

%"However, extending test-time compute does not lead to constant performance improvement in today’s reasoning LLMs, where accuracy improvements diminish and eventually stagnate after a certain number of decoding steps. This has fueled discussions around “LLM overthinking” (Ghosal et al., 2025; Chen et al., 2025b; Fan et al., 2025; Cuadron et al., 2025; Li et al., 2025), where models expend excessive computation on problems, with the additional reasoning steps yielding minimal or no benefit to the final answer." -- Parathinker

%"We systematically investigate how the marginal utility of additional reasoning tokens changes as compute budgets increase. We find that marginal returns diminish substantially at higher budgets and that models exhibit “overthinking” -- When More Thinking Hurts: Overthinking in LLM Test-Time Compute Scaling

We assume that the model $H$ is more capable (in the limit) and also costs more per token.
\[
M_H > M_L \, \quad c_H > c_L
\]

For a given model $m$ and reward share $\alpha$, the Agent solves:
\[
  \max_{x \geq 0}\; U(\alpha,m,x) \implies \max_{x \geq 0}\; \alpha M_m\!\left(1 - e^{-k_m x}\right) - c_m x.
\]

The FOC gives:
\[
  \alpha M_m k_m e^{-k_m x^*} = c_m.
\]

Solving for the optimal effort for model $m$, $x^*_m$:
\begin{align}
    x^*_m = \frac{1}{k_m}\ln\!\left(\frac{\alpha M_m k_m}{c_m}\right). \label{eq:opt_effort}
\end{align} 

This is valid (i.e.\ $x^*_m \geq 0$) only when $\frac{\alpha M_m k_m}{c_m} \geq 1$. 
\textbf{Activation threshold $\tau_m$} is defined as the minimum contract share $\alpha$ at which it is viable for the Agent to use model $m$ to start producing tokens. In other words, for any reward share less than $\tau_m$, the marginal cost exceeds marginal benefit for model $m$. It is given by:
\[
  \tau_m = \frac{c_m}{M_m k_m}.
\]

\begin{itemize}
  \item If $\alpha \leq \tau_m$: the Agent will not spend any tokens on reasoning, \emph{i.e.}, $x^*_m = 0$ and $U_A = 0$.
  \item If $\alpha > \tau_m$:  $x^*_m = \frac{1}{k_m}\ln\!\left(\frac{\alpha}{\tau_m}\right)$.
\end{itemize}

Substituting $x^*_m$ into Eq.~\ref{eq:utilities}, and for simplicity, we denote $U^*(\alpha, m, x^*_m)$ as $U_m(\alpha)$ which is given by:

\begin{align}
    U_m(\alpha) = M_m\!\left[\alpha - \tau_m - \tau_m\ln\!\left(\frac{\alpha}{\tau_m}\right)\right], \quad \alpha > \tau_m. \label{eq:opt_util}
\end{align}
  
Note that $U_m(\tau_m) = M_m[\tau_m - \tau_m - \tau_m \cdot 0] = 0$. So utility is continuous at the threshold.
%and the agent is exactly indifferent between participating and not at $\alpha = \tau_m$. 
The first derivative is:
\[
    U'_m(\alpha) = M_m \!\left[ 1- \frac{\tau_m}{\alpha}\right]
\]
It is positive when $\alpha > \tau_m$, and the second derivative is also positive, which implies $U_m(\alpha)$ is convex and increasing.
%\kate{It is worthwhile to add a sentence explicitly stating what you have shown for the agent. You never want your reader to be thinking hard about the implications of anything.}
We have now derived the optimal token budget ($x^*_m$) and resulting utility $U_m(\alpha)$ for the Agent using any model $m$ given a linear contract $\alpha$ proposed by the Principal. 

\subsection{Switching Threshold $\theta$}
Next, we characterize the optimal model choice of the Agent given the linear contract $\alpha$. Specifically, at what value of $\alpha = \theta$ the Agent's optimal model choice switches from one model to another. We consider two scenarios depending on which model activates first, \emph{i.e.}, which model requires the least reward share $\alpha$ to start producing tokens.%\kate{You use terms like "viable" and "activate" but it is never 100\% clear what you mean until later.}

\begin{comment}
\begin{figure}[h]
    \centering
    \includegraphics[width=0.8\columnwidth]{Figures/Standard_Order_with_Differential.pdf}
    \caption{Standard Ordering. Agent's utilities $U_m(\alpha)$ under optimal token budget, illustrating the Agent's optimal model choice in different intervals over the contract parameter $\alpha$.\kate{Remove or put into an appendix} [can REMOVE lower panel if you think it is not really useful]}
    \label{fig:standard_order_illustration}
\end{figure}
\end{comment}

\begin{figure*}[h]
    \centering
    \begin{subfigure}[b]{0.48\textwidth}
        \includegraphics[width=\textwidth]{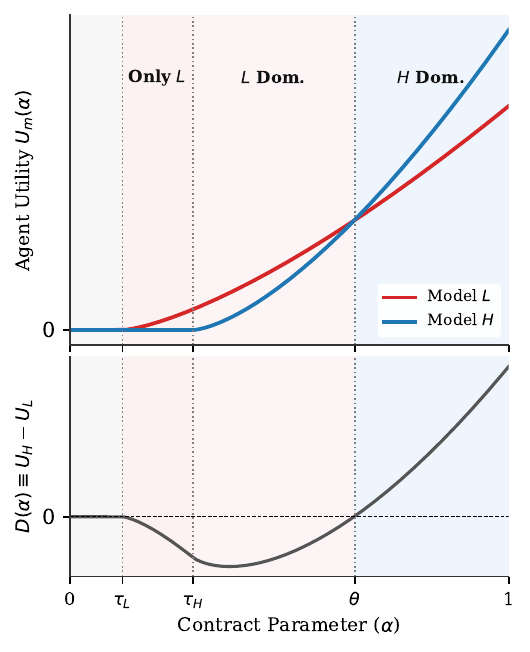}
        \caption{Standard Order $\tau_L < \tau_H$}
        \label{fig:standard_order_illustration}
    \end{subfigure}
    \hfill
    \begin{subfigure}[b]{0.48\textwidth}
        \includegraphics[width=\textwidth]{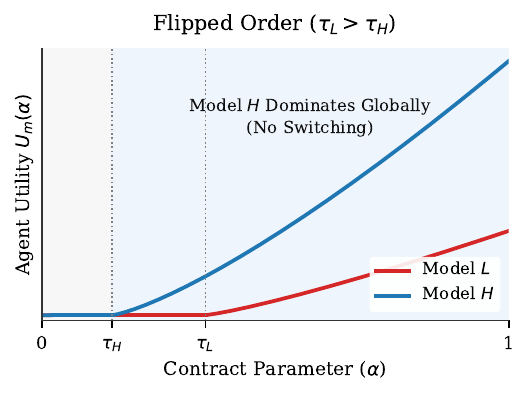}
        \caption{Flipped Order $\tau_L > \tau_H$}
        \label{fig:flipped_order_illustration}
    \end{subfigure}
    \caption{Illustration of model utilities for Agent $U_m(\alpha)$ under both scenarios.}
\end{figure*}

\subsubsection{Standard Order $\tau_L < \tau_H$}
Model $L$ activates earlier than $H$, and therefore, at $\tau_H$,  $U_L(\tau_H) > 0$, whereas model $H$ is just activated, \emph{i.e.}, $U_H(\tau_H)$ = 0. 

We define the utility differential as:
\begin{equation}
D(\alpha) \equiv U_H(\alpha) - U_L(\alpha)
\label{eq:util_diff}
\end{equation}

Differentiating $D(\alpha)$ yields:
\[
D'(\alpha) = M_H - M_L - \frac{M_H\tau_H - M_L \tau_L}{\alpha} 
\]

\[
D''(\alpha) = \frac{M_H \tau_H - M_L \tau_L}{\alpha^2}
\]
Under standard order ($\tau_H > \tau_L$) and the fact that $M_H > M_L$, the second derivative $D''$ is positive throughout. Therefore, the utility differential $D(\alpha)$ is convex. At the threshold, $D(\tau_H) = -U_L(\tau_H) < 0$. $D'$ is also negative at this point. Due to convexity and the negative slope, as we increase $\alpha$ beyond $\tau_H$, $D$ decreases further until $D' = 0$ and then begins to increase. Therefore, there is exactly one solution $\theta$ in $[\tau_H, \infty)$ where $D(\theta) = 0$ (see Fig.~\ref{fig:standard_order_illustration}). The switching threshold $\theta$ exists within the permissible range $[0,1]$ only if $D(1)$ is positive. Otherwise, model $L$ will dominate throughout the domain. 

%In this case, both models are active for $\alpha > \tau_H$. However, the Agent chooses $L$ until $\alpha = \theta$, and then switches over to model $H$ beyond. 

If $D(1) > 0$, the domain of $\alpha$ is partitioned into 4 regions (shown in Fig.~\ref{fig:standard_order_illustration} in Appendix A). Neither model is viable in the first interval (until $\tau_L$) and only $L$ is viable in the second. Both models are viable in the third region where $\alpha > \tau_H$, but $L$ is still more rewarding for the Agent. Beyond $\theta$, the quality `premium' of $H$ is worth its higher cost, and the Agent's model choice jumps from $L$ to $H$. Note that the optimal token budget is discontinuous at this point (Eq.~\eqref{eq:opt_effort}).

\subsubsection{Flipped Order $\tau_L > \tau_H$}
In this case, model $L$ activates at a higher contract share than the higher model $H$. Because $\tau_H < \tau_L$, it follows that $\frac{\tau_H}{\alpha} < \frac{\tau_L}{\alpha}$, which implies:
\[
\left(1 - \frac{\tau_H}{\alpha}\right) > \left(1 - \frac{\tau_L}{\alpha}\right)
\]

Coupled with the fact that $M_H > M_L$, this implies $U'_H(\alpha) > U'_L(\alpha)$ $\forall \alpha > \tau_L$.  Since Model $H$ activates earlier and climbs strictly faster at every point, model $H$ dominates $L$ throughout the domain, and there is no switching in this case (see Fig.~\ref{fig:flipped_order_illustration} for illustration).

%\kate{Like I said before, it would be nice to tell the reader what they need to take away from the analysis. That is, given some $\alpha$ you have derived the optimal action for the Agent.}

We have thus derived the optimal token budget (Eq.~\eqref{eq:opt_effort}) for each model $m$ given a contract from the Principal, parametrized by $\alpha$. Using that, we then derived the optimal model choice for the Agent, characterized by the switching threshold $\theta$ (its existence and meaning determined by the two scenarios outlined above).

\subsection{Principal's Optimization}
\label{sec:principal_optim}
%\kate{Now say that you will figure out what the principal will do, given you have characterized the Agents best-response.}

Using the two-dimensional best response of the Agent for any given contract $\alpha$, we now derive the Principal's optimal linear contract. The Principal maximises $V(\alpha,m,x) = (1-\alpha)\cdot q_m(x^*_m(\alpha))$ over $\alpha$, but $m$ here is not a free choice for the Principal; rather it is determined by the Agent's best response. For a given model $m$, Agent's optimal token budget $x^*_m$ is given by~\eqref{eq:opt_effort}. The corresponding quality is $q_m(x^*_m) = M_m(1-\tau_m/\alpha)$ and therefore, the Principal's utility for a given model is:
\[
  V_m(\alpha) = (1-\alpha)\, q_m(x^*_m) 
  = M_m\!\left[1+\tau_m - \alpha - \frac{\tau_m}{\alpha}\right]
\]
This is the payoff the Principal would get \emph{if} the Agent's optimal model choice would be $m$ at share $\alpha$. Differentiating,
\[
  \frac{dV}{d\alpha} = M_m\!\left(\frac{\tau_m}{\alpha^2} - 1\right) = 0
  \;\Longrightarrow\; \alpha^*_m = \sqrt{\tau_m},
\]

\[\dfrac{d^2V}{d\alpha^2} = -\dfrac{2M_m\tau_m}{\alpha^3} < 0\]

So $\alpha^*_m$ is a maximum of $V_m$ over $\alpha>0$, and $V_m$ is strictly concave, single-peaked at $\sqrt{\tau_m}$. We assume $\tau_m<1$ for both models, so that $\sqrt{\tau_m}\in(\tau_m,1)$ is a feasible share.

\subsubsection{Standard order $\tau_L<\tau_H$.}
If the switching threshold $\theta$ exists within $[0,1]$, then the Agent picks $L$ on $[\tau_L,\theta)$ and $H$ on $[\theta,1]$. $V_m(\alpha)$ is achievable only when restricted to the interval where $m$ is actually the
Agent's choice, and because each $V_m(\alpha)$ is single-peaked, the constrained optimum on each interval is either the peak (if it falls inside the interval) or the boundary (if the peak falls outside), i.e., clip $\alpha^*_m$ such that $m$ is chosen by the Agent.

\begin{equation}
  \alpha_L^\dagger = \min\!\big(\sqrt{\tau_L},\, \theta\big), \qquad
  \alpha_H^\dagger = \max\!\big(\sqrt{\tau_H},\, \theta\big).
  \label{eq:alpha_dagger}
\end{equation}

If $\sqrt{\tau_L} \ge \theta$, the function $V_L(\alpha)$ is increasing in $[\tau_L,\theta)$ with the peak not yet attained. So, the maximum value is at the boundary, but note that the Agent switches to model $H$ at $\theta$. Therefore, the maximum value of $V_L$ is the left-hand limit of $V_L(\theta)$. Symmetrically, if $\sqrt{\tau_H} \le \theta$, $V_H(\alpha)$ is decreasing on $[\theta,1]$ and its maximum is attained at the left endpoint $\theta$. 

%Note that $V$ is generally \emph{discontinuous} at $\theta$: the agent's utility is continuous there by definition of $\theta$, but $q_L(x_L^*(\theta))$ and $q_H(x_H^*(\theta))$ need not coincide, so $U_P(\theta,L^-)$ and $U_P(\theta,H)$ differ in general. This is why $\theta$ itself — not just the two interior optima — must be checked as a candidate.

The Principal's optimal contract is then
\begin{equation}
    \alpha^* = \arg\max_{\alpha\in\{\alpha_L^\dagger,\ \alpha_H^\dagger\}} V(\alpha, m(\alpha)) 
    \label{eq:principals_opt}
\end{equation}
  
Note that if $D(1)\le 0$, \emph{i.e.}, the switching $\theta$ does not exist in the domain, $L$ dominates throughout $[0,1]$ and the Principal's problem reduces to $\alpha^*=\sqrt{\tau_L}$.

\paragraph{Flipped order $\tau_L>\tau_H$.}
Here $H$ weakly dominates $L$ everywhere it is active, so the Agent never chooses $L$ and the Principal's problem reduces to the single unconstrained optimization $\alpha^* = \sqrt{\tau_H}$.

Thus, we have derived the optimal linear contract for the Principal and characterized the Agent's best response. Further derivation of the first-best benchmark, total surplus, and agency costs are in Appendix A.

\subsection{Burn-in Tokens for Reasoning Models}
\label{sec:burn_in}
The modeling choice of saturation function for the LLM quality is appropriate for simpler tasks with instruction-tuned models. For complex tasks, especially with reasoning models, there is a burn-in period, where accuracy is zero despite spending tokens. We denote $\tilde{x}$ to be the raw tokens spent, and $b_m$ denotes the number of burn-in tokens for model $m$. Then, 

\begin{align}
    q_m(\tilde{x}) = \begin{cases}
        0, & \tilde{x} < b_m \\
        M_m\!\left(1 - e^{-k_m (\tilde{x}-b_m)}\right), & \tilde{x} \geq b_m
    \end{cases}      
    \label{eq:saturation_fn_burnin}
\end{align}

Rewriting $x = \tilde{x} - b_m$ as the effective tokens spent by the Agent, we get the basic saturation function from earlier~\eqref{eq:saturation_func}. However, the associated cost $c_m b_m$ was not considered. The true utility of the Agent is:
\[
\mathcal{U}(\alpha,m,x) = \alpha \cdot q_m(x) - c_m x - c_m b_m
\]

However, this extra fixed cost is a constant in terms of $x$, and therefore the optimal effective token $x^*_m$ does not change from~\eqref{eq:opt_effort}. Note that the optimal true token $\tilde{x}^*_m$ is shifted by $b_m$. The Agent utility under a given model $m$ now becomes:
\[
\mathcal{U}_m(\alpha) = M_m\!\left[\alpha - \tau_m - \tau_m\ln\!\left(\frac{\alpha}{\tau_m}\right)\right] - c_m b_m 
\]

Note that the burn-in utility is simply the same $U_m(\alpha)$ as before with an additional negative term corresponding to the fixed cost due to the burn-in tokens. While the activation threshold $\tau_m$ ensured the Agent's participation in the base case, now the Agent's utility is in fact negative at $\tau_m$ due to the (fixed) burn-in cost. The \emph{participation threshold} $\tau^P_m$ defines the lowest value of $\alpha$ such that the Agent breaks even using model $m$, \emph{i.e.}, $\mathcal{U}_m(\tau^P_m) = 0$. 
\[
M_m \left[ \tau^P_m - \tau_m - \tau_m \ln\left(\frac{\tau^P_m}{\tau_m}\right) \right] = c_m b_m
\]

Simplifying, we get
\[
\left(\frac{\tau^P_m}{\tau_m}\right) - 1 - \ln\left(\frac{\tau^P_m}{\tau_m}\right) = k_m b_m
\]

Rewriting the improper fraction as $z$, consider $f(z) = z - 1 - \ln z$ evaluated over the active domain $z > 1$. Computing its derivatives:
\[
f'(z) = 1 - \frac{1}{z} > 0, \quad f''(z) = \frac{1}{z^2} > 0
\]
Thus, $f(z)$ is strictly increasing and strictly convex. Its inverse function $f^{-1}(x)$ is strictly increasing and strictly concave. This allows an exact closed-form expression of the participation threshold:
\begin{equation}
\tau^P_m = \tau_m \cdot f^{-1}(k_m b_m)
\label{eq:participation_threshold}
\end{equation}

%Since the domain of $f(z)$ is $z > 1$, we can see that the participation threshold is a rightward shift of the activation threshold. 

\subsubsection{Switching Threshold $\theta_0$}

We rewrite the burn-in utility as $\mathcal{U}_m(\alpha) = U_m(\alpha) - c_m b_m$, and reuse the baseline differential $D(\alpha) \equiv U_H(\alpha) - U_L(\alpha)$ from~\eqref{eq:util_diff}. The burn-in switching threshold $\theta_0$ satisfies $\mathcal{U}_H(\theta_0) =\mathcal{U}_L(\theta_0)$, i.e.\ $U_H(\theta_0) - c_H b_H = U_L(\theta_0) - c_L b_L$. Rearranging:

\begin{equation}
    D(\theta_0) = c_H b_H - c_L b_L
    \label{eq:burnin_eq_gen}
\end{equation}

This is not closed-form in general: $D$ is a transcendental function of $\alpha$. In practice, $\theta_0$ is found numerically as the root of~\eqref{eq:burnin_eq_gen}, via bracketed root-finding like Brent's method.

To obtain sharper qualitative results, we restrict to a special case with common $k$ and $b$ for both models. Under this assumption, $\theta_0$ is a strict rightward shift of the base-model threshold $\theta$ (i.e., burn-in delays switching) in the Standard Order case ($\tau_L < \tau_H$). In the Flipped Order case ($\tau_H < \tau_L$), $H$ continues to dominate everywhere, exactly as in the base model. See Appendix B for further details.

\section{Experimental Setup}
We evaluate our proposed framework empirically across two distinct domains: advanced mathematical reasoning and multi-discipline question answering. Specifically, we evaluate model pairings on the \texttt{MATH} dataset (difficulty levels 3 and 4) and the \texttt{MMLU-Pro} dataset~\cite{hendrycksmath2021, wang2024mmlu}. Our model suite consists of instruction-tuned models (e.g., Llama-3.2), distilled reasoning models (e.g. DeepSeek-R1-Distill-Qwen), and recent edge reasoning models (e.g. Gemma4). We consider models of different sizes and consider pairings of same and different types. In this section, we describe how we calibrate the production function and then describe the learning processes.

\subsection{Calibration of Production Functions}
\label{sec:calibration}
\begin{figure}[h]
    \centering
    \includegraphics[width=0.75\columnwidth]{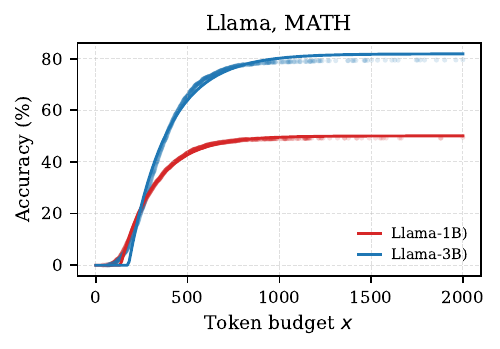}
    \caption{Accuracy \textit{vs.} budget for Llama models on \texttt{MATH}.}
    \label{fig:llama_calib}
\end{figure}

To bridge our analytical model with real-world LLM performance, we first empirically calibrate the production parameters $(M_m, k_m, b_m)$ for each model on both evaluation datasets. For each dataset and model, we let the model answer every question under a maximum token budget of $x_{\max}$ tokens at temperature $0$. Let $x_i$ denote the number of tokens used to answer question $i$. For a dense grid of budgets from $0$ to $x_{\max}$, accuracy at budget $x$ is the percentage of questions answered correctly using at most $x$ tokens, i.e.\ with $x_i \le x$ and a correct answer. This produces an empirical production curve for each model-dataset pair, to which we fit the saturating function with burn-in tokens (Eq.~\eqref{eq:saturation_fn_burnin}) via nonlinear least squares (using \texttt{curve\_fit} method from \texttt{scipy}). Table~\ref{tab:math_calibration} reports the fitted parameters, along with $R^2$, for both tasks. Figure~\ref{fig:llama_calib} shows the accuracy at each budget level along with the fitted curves for the Llama models in MATH (refer Fig.~\ref{fig:cal_matrix} in Appendix E for all other models and tasks).

\begin{table}[h]
\centering
\small
\setlength{\tabcolsep}{4pt}
\begin{tabular}{lccccc}
\toprule
\textbf{Model} & $M_m$ & $k_m$ & $b_m$ & $R^2$ & \textit{RMSE} \\
\midrule
\multicolumn{6}{c}{\textit{MATH task domain}} \\
\midrule
Llama-3.2-1B-Instruct       & 50 & 0.00529 & 134 & 0.998 & 0.772 \\
Llama-3.2-3B-Instruct       & 82 & 0.00473 & 177 & 0.994 & 2.153 \\
\midrule
DS-R1-Distill-1.5B    & 75 & 0.00081 & 982 & 0.998 & 0.979 \\
DS-R1-Distill-7B      & 92 & 0.00073 & 1009 & 0.997 & 1.373 \\
\midrule
Gemma-4-E2B-it              & 84 & 0.00159 & 430 & 0.999 & 0.952 \\
Gemma-4-E4B-it              & 85 & 0.00190 & 382 & 0.997 & 1.411 \\
\midrule
\multicolumn{6}{c}{\textit{MMLUPro task 
domain}} \\
\midrule
Llama-3.2-1B-Instruct       & 22 & 0.00366 & 162 & 0.983 & 0.928 \\
Llama-3.2-3B-Instruct       & 40 & 0.00474 & 164 & 0.986 & 1.589 \\
\midrule
DS-R1-Distill-1.5B    & 24 & 0.00212 & 479 & 0.996 & 0.536 \\
DS-R1-Distill-7B      & 41 & 0.00141 & 440 & 0.999 & 0.414 \\
\midrule
Gemma-4-E2B-it              & 59 & 0.00202 & 422 & 0.997 & 1.029 \\
Gemma-4-E4B-it              & 68 & 0.00235 & 413 & 0.998 & 1.084 \\
\bottomrule
\end{tabular}
\caption{Model calibration and fitted params for both tasks.}
\label{tab:math_calibration}
\end{table}

\begin{comment}
\begin{table}[htbp]
\centering
\small
\begin{tabular}{lccccc}
\toprule
\textbf{Model} & $M$ & $k$ & $x_0$ & $R^2$ & \textit{RMSE} \\
\midrule
Llama-3.2-1B-Instruct       & 22 & 0.00366 & 162 & 0.983 & 0.928 \\
Llama-3.2-3B-Instruct       & 40 & 0.00474 & 164 & 0.986 & 1.589 \\
\midrule
DS-R1-Distill-1.5B    & 24 & 0.00212 & 479 & 0.996 & 0.536 \\
DS-R1-Distill-7B      & 41 & 0.00141 & 440 & 0.999 & 0.414 \\
\midrule
Gemma-4-E2B-it              & 59 & 0.00202 & 422 & 0.997 & 1.029 \\
Gemma-4-E4B-it              & 68 & 0.00235 & 413 & 0.998 & 1.084 \\
\bottomrule
\end{tabular}
\caption{Model Calibration and Saturation Function Fit Parameters for the \textbf{MMLUPro} Domain. [shift to Appendix?]}
\label{tab:mmlu_calibration}
\end{table}
\end{comment}

%To formalize the economic environment, we normalize the token cost of the lower-tier model to $c_L = 0.005$ monetary units per token. The token cost of the frontier model, $c_H$, is systematically scaled as a free parameter relative to $c_L$ to explore interesting scenarios of switching between models.

We anchor the token cost of the lower-tier model to $c_L = 0.005$ monetary units per token in every pairing. This fixes a common scale and the variation between various pairings is through the capability parameters $(M_m,k_m,b_m)$ and the cost ratio $c_H/c_L$. The $H$ model's cost $c_H$ is chosen per pairing: for most pairings we select $c_H$ to fall within the range that yields Standard Order with a genuine switching threshold $\theta_0 \in (0,1]$, giving representative cases of model switching; for a small number of pairings we deliberately choose $c_H$ outside this range to illustrate the Flipped Order and always dominant $L$ scenarios discussed in the Model Section.

\subsection{Learning Processes of Agent and Principal}
\label{sec:linucb}
To evaluate how efficiently the Agent can learn the optimal mechanism without prior knowledge of the calibration parameters, we frame the model choice and budget selection problem as a Contextual Multi-Armed Bandit. At each sequential round $t$, the Agent receives a contract stake $\alpha_t$ drawn uniformly at random from the domain $\alpha_t \in [0, 1]$, which acts as the context. The task at each round $t$ is a randomly sampled set of $16$ questions from the dataset. See Appendix E for further experimental details and parameters. 

The action space $\mathcal{A}$ is structured as a joint choice space $\mathcal{A} = \mathcal{M} \times \mathcal{X}$, where $\mathcal{M} = \{L, H\}$ represents model choices and $\mathcal{X} = \{x^{(1)}, x^{(2)}, \dots, x^{(N)}\}$ is a discretized set of token budgets spanning from $x_{min}$ up to $x_{max}$ tokens. For a given context $\alpha_t$, the net reward observed by choosing arm $a = (m, x)$ is given by $R_t(a) = \alpha_t \cdot y_{t,m}(x) - c_m \cdot x$
where $y_{t,m}(x)$ is the empirical accuracy achieved by model $m$ under budget $x$ (questions answered correctly by the LLM). 

We deploy the LinUCB algorithm~\cite{li_contextual-bandit_2010} to model the expected reward of each arm.
Crucially, 
%while the true expected accuracy curve is a saturating exponential,
the optimal utility of the Agent $U_m(\alpha)$ maps exactly to a linear combination of $\alpha$ and $\ln(\alpha)$ due to the structure of the Agent's FOC; see Eq.~\eqref{eq:opt_util}. To ensure faster convergence\footnote{We also explored using standard UCB with discretized $\alpha$; similar trends (Fig.~\ref{fig:simple_ucb} in App. E) but takes longer to converge.}, we construct a handcrafted context feature vector: $\begin{bmatrix} 1 & \alpha_t & \ln(\alpha_t) \end{bmatrix}$.

%[\boldsymbol{\phi}(\alpha_t) = \begin{bmatrix} 1 & \alpha_t & \ln(\alpha_t) \end{bmatrix}\]

%LinUCB is a contextual bandit algorithm~\cite{li_contextual-bandit_2010}: it assumes each arm's expected reward is a linear function of the current context, and maintains a running estimate of the weights (using ridge-regression) from observed rewards. Similar to UCB, at each round, it adds an exploration bonus which shrinks as more data accumulates for that arm. We use a smaller exploration weight at the model-choice level ($\gamma=1$) and a larger one at the token-budget level ($\gamma=2$): since the model-choice level's reward estimate depends on the token-budget policy already being close to its optimum, we push the token-budget level to converge faster, so that by the time the model-choice level settles, it is learning on reliable, near-optimal reward signals.

%We simulate this loop over a horizon of $T = 2,000$ episodes per model pairing to observe how the learned policies compare against our analytical thresholds.

The Principal's learning is modeled as classic (non-contextual) UCB, with arms representing a discretized grid of contract shares $\alpha \in [0,1]$. Unlike the Agent, the Principal has no natural context to condition on as it chooses $\alpha$ rather than responding to it. Each round, the Principal selects an arm (contract share), the frozen, previously-trained Agent best-responds with its model and token-budget choice, and the Principal observes its realized payoff; $(1-\alpha_t) \cdot y_{t,m}(x)$. This reward is used to update the estimate for the pulled arm.

\section{Results}
In this section, we compare the learned policies against our theoretical model with fitted parameters across different model pairings and two task domains. 

\subsection{\texttt{MATH} domain}

\begin{figure}[H]
    \centering
    \includegraphics[width=0.75\columnwidth]{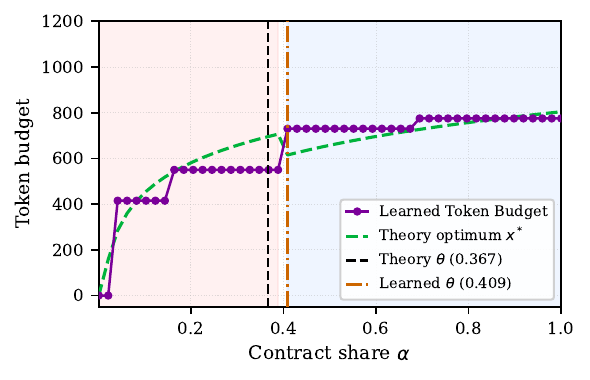}
    \caption{MATH: Llama 1B vs. Llama 3B}        
    \label{fig:math_llama1b_vs_llama3b_main}
\end{figure}

First, we evaluate our model on the MATH domain~\cite{hendrycksmath2021}. The LLM prompts for question-answering are included in App. G. The theoretical value of $\theta_0$ is calculated using the calibrated parameters in Table~\ref{tab:math_calibration} and solving for the root of~\eqref{eq:burnin_eq_gen} using \texttt{brentq} method in \texttt{scipy} package. We train the LinUCB controller for $2,000$ episodes, the learned policy is used to greedily choose model and token budget for 50 equally interspersed values of $\alpha \in [0.001,1]$ to obtain the learned value of $\theta_0$. These numbers are reported in Table~\ref{tab:math_res}. We observe that the learned values are generally close to the estimated values (typically within about $10-15\%$); 
%the difference is to be expected due to the discretized budget space and noisy LLM inference. 
the difference is an artifact of mapping the continuous budget $x$ onto a discrete token budget space ($N=21$ bins) compounded by noisy LLM inference. Fig.~\ref{fig:math_llama1b_vs_llama3b_main} shows the learned token budget along with the learned switching threshold for the Llama 1B \textit{vs.} 3B pairing. Similar figures for all other pairings are in App. E.

\begin{table}[h]
\centering
%\small
\begin{tabular}{lccc}
\toprule
\textbf{Model Pairing} & $c_H$ & $\theta_0$ (est.) & $\theta_0$ (learn) \\
\midrule
\multicolumn{4}{c}{\textit{Intra-Family Pairs}} \\
\midrule
Llama 3.2: 1B \textit{vs.} 3B   & 0.0200 & 0.367 & 0.409 \\
DeepSeek R1: 1.5B \textit{vs.} 7B & 0.0075 & 0.731 & 0.817\\
Gemma 4: E2B \textit{vs.} E4B   & 0.0060 & 0.255 & 0.286 \\
\midrule
\multicolumn{4}{c}{\textit{Inter-Family Pairs}} \\
\midrule
Llama 1B \textit{vs.} Gemma 4B  & 0.0100 & 0.419 & 0.388 \\
Llama 1B \textit{vs.} DS 1.5B   & 0.0075 & 1.331 & $L$ dom. \\
DS 1.5B \textit{vs.} Gemma 4B   & 0.0075 & \text{N/A} & $H$ dom. \\
\bottomrule
\end{tabular}
\caption{\textbf{MATH Domain}: Learned model choices $(\theta_0)$ for the LinUCB controller. We fix $c_L = 0.005$ for all configurations.}
\label{tab:math_res}
\end{table}

While values of $c_H$ were mostly chosen to induce standard order model switching, we also included some other scenarios. Llama 1B \textit{vs.} DS 1.5B is one such case where the switching threshold $\theta_0$ lies outside the domain $[0,1]$. Intuitively, $H$, which is the Deepseek model here, is significantly more capable than Llama 1B (about 50\%; see Table~\ref{tab:math_calibration}). However, Deepseek has higher burn-in tokens and that combined with the higher cost makes it unattractive for any $\alpha$ within the domain. The LinUCB Agent learns to always pick $L$ in this scenario. 

The cost $c_H$ for the pairing DS 1.5B \textit{vs.} Gemma 4B was chosen to exhibit yet another interesting scenario. Here, $\tau_L = 0.0833$ and $\tau_H = 0.0464$. That is, this specific configuration belongs to the \textit{flipped order} scenario, where $H$ activates earlier and dominates $L$ everywhere. There is no switching threshold $\theta_0$ in this scenario and the LinUCB Agent matches this exactly, where $H$, the Gemma 4B model dominates everywhere. This is explained by the fact that Gemma models have higher capability ceiling than Deepseek R1 1.5B on this task and they have lower burn-in tokens (see calibration Table~\ref{tab:math_calibration}). So, despite the slightly higher cost, it is economically rational to always use the expensive model.

Instead of the LinUCB bandit controller, we also explored prompting a much stronger LLM to choose the model and token budget in Appendix F; results are broadly consistent with optimal model choice but not the optimal budget. 

\subsection{\texttt{MMLUPro} Domain}
\begin{figure}[h]
    \centering
    \includegraphics[width=0.75\columnwidth]{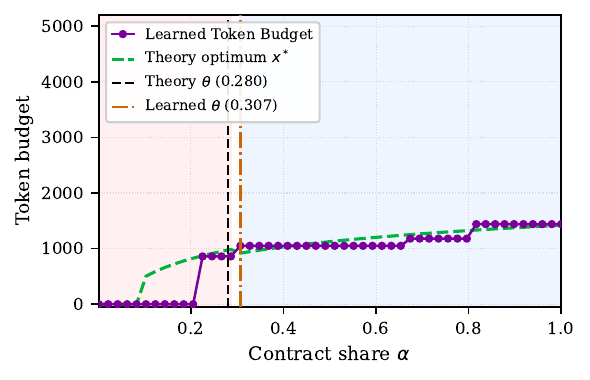}        \caption{MMLUPro: DeepSeek 1.5B vs. Gemma 4B}        \label{fig:ds1_5b_vs_gemma4b_main}
\end{figure}

We also evaluate our results on the MMLUPro domain~\cite{wang2024mmlu}, and the results are shown in Table~\ref{tab:mmlu_res}. Similar to the results from the MATH domain, the bandit algorithms learned the switching thresholds that are generally close to the estimated theoretical value of $\theta_0$. Figure~\ref{fig:ds1_5b_vs_gemma4b_main} shows the learned token budget and model choice for the DeepSeek 1.5B vs. Gemma 4B pairing (other pairings are in Fig.~\ref{fig:mmlu_policy_matrix} in App. E).

\begin{table}[h]
\centering
%\small
\begin{tabular}{lccc}
\toprule
\textbf{Model Pairing} & $c_H$ & $\theta_0$ (est.) & $\theta_0$ (learn) \\
\midrule
\multicolumn{4}{c}{\textit{Intra-Family Pairs}} \\
\midrule
Llama 3.2: 1B \textit{vs.} 3B   & 0.0200 & 0.589 & 0.633 \\
DeepSeek R1: 1.5B \textit{vs.} 7B & 0.0075 & 0.309 & 0.286\\
Gemma 4: E2B \textit{vs.} E4B   & 0.0060 & 0.339 & 0.286 \\
\midrule
\multicolumn{4}{c}{\textit{Inter-Family Pairs}} \\
\midrule
Llama 1B \textit{vs.} Gemma 4B  & 0.0200 & 0.520 & 0.592 \\
Llama 1B \textit{vs.} DS 1.5B   & 0.0075 & 6.931 & $L$ dom. \\
DS 1.5B \textit{vs.} Gemma 4B   & 0.015 & 0.410 & 0.409 \\
DS 1.5B \textit{vs.} Gemma 2B   & 0.015 & 0.280 & 0.307 \\
DS 1.5B \textit{vs.} Llama 3B   & 0.006 & \text{N/A} & $H$ dom. \\
\bottomrule
\end{tabular}
\caption{\textbf{MMLUPro Domain}: Learned model choices (determined by $\theta_0$). We fix $c_L = 0.005$ for all pairings.}
\label{tab:mmlu_res}
\end{table}

\FloatBarrier
\subsection{Principal's Learning}
In this section, we show that the Principal can learn the appropriate linear contract when interacting with a trained best-responding Agent from previous sections. The Principal's learning process is outlined in the previous section. 

\begin{figure}[h]
    \centering
    \includegraphics[width=0.75\columnwidth]{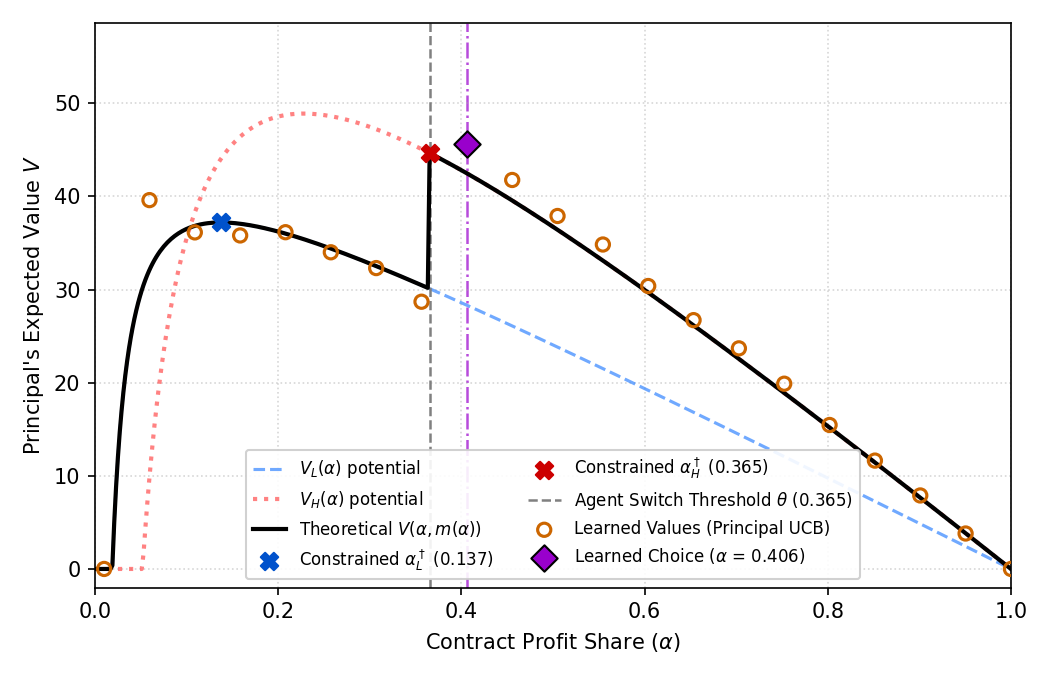}
    \caption{Principal's Learning for Llama 3.2: 1B \textit{vs.} 3B}
    \label{fig:llama_principal}
\end{figure}

We use $21$ arms for the Principal and train it for $3,000$ episodes against a frozen, trained Agent from the previous section. We test this in the \texttt{MATH} domain with different model pairings. For the Llama 3.2: 1B \textit{vs.} 3B pairing, Fig.~\ref{fig:llama_principal} shows the theoretically expected utility values of the Principal over the domain (in black) and the learned value at each arm (orange circles). The constrained $\alpha^\dagger_m$ (see Eq.~\eqref{eq:alpha_dagger}) for each model are calculated and shown based on the calibrated parameters. Finally, the arm (contract share) chosen by a greedy Principal post training is shown as the purple diamond, which is close to theoretical expectations. Similar results are seen for the other pairing; refer to Fig.~\ref{fig:principal_matrix} in App. E.

\begin{comment}
\begin{figure}[t]
    \centering
    \includegraphics[width=\columnwidth]{Figures/gemma.png}
    \caption{Principal's Learning for Gemma 4: E2B \textit{vs.} E4B.}
    \label{fig:gemma_principal}
\end{figure}
\end{comment}

\FloatBarrier
\section{Conclusion}

In this paper, we extended the classical Principal-Agent framework to address the emerging moral hazard in autonomous LLM delegation, a setting where an Agent dynamically selects both a reasoning technology (model choice) and a continuous effort level (token budget). By modelling the LLM inference process as a saturating production function, we derived the Agent's optimal token allocation and characterized the switching threshold at which it becomes economically rational to adopt a more capable, yet more expensive, model. Furthermore, we determined the Principal's optimal linear contract that maximizes expected utility despite the Agent's hidden actions. Our empirical calibrations on the MATH and MMLUPro benchmarks, paired with simulations using contextual bandit algorithms, demonstrated that both the Principal and the Agent converge toward strategies consistent with our theoretical equilibrium.

As LLM inference-time scaling and budget-aware reasoning continue to advance, the economic implications of optimizing joint model and budget choices will only grow in relevance. While our current framework assumes verification is costless, we have established the foundations for constant-cost verification in Appendix D, and it would be worthwhile to explore other, more complex verification cost families in the future. Additionally, extending this framework to non-ground truth settings presents a compelling direction for future research. In such environments, both the evaluation of correctness and the valuation of the task depend entirely on the Principal's specific type, introducing severe information asymmetry. Without knowing the exact type of the Principal, the Agent's best response would require reasoning over the type distribution, bridging mechanism design with Bayesian delegation in complex agentic markets.

%Future work: Recent advances in optimizing budgets bodes well for our economic model of considering model choices along with budget levels; making it even more relevant. We assumed verification is possible and free, we explore constant cost verification in Appendix B but it would be worthwhile exploring other cost families. In non-groundtruth settings, where both evaluation (how correct it is) and valuation (how important it is) depend on the Principal's type; we would then have different valuation functions $v_t(q(x))$ for type $t$. This introduces information asymmetry to the problem, and without knowing the exact type of the Principal, the Agent's best response would need to reason over the type distribution.
\clearpage
\bibliographystyle{plain}
\bibliography{refs}

\clearpage
\appendix
\onecolumn

\begin{center}
    \section*{Appendix}
\end{center}
\section{Surplus and Agency Costs}
\label{sec:first-best-surplus}

We now consider the first-best benchmark for the base model. In the first-best scenario, a single party holds the model, reaps the benefit and spends compute cost, \emph{i.e.}, single party receives the total social surplus, with no moral hazard. This benchmark is the limiting case $\alpha = 1$ of the contract space above. The optimal token budget is the same as optimal effort for model $m$ at $\alpha=1$ from Eq.~\eqref{eq:opt_effort}: 
\[
x^{*,FB}_m = \frac{1}{k} \, \ln\!\left(\frac{1}{\tau_m}\right) = x^*_m(1)
\]
At $\alpha=1$ the Agent's payoff $U_m(1) = q_m(x_m^*(1)) - c_m x_m^*(1)$ equals total surplus under model $m$ at efficient effort, so $D(1) = U_H(1) - U_L(1)$ is exactly the comparison of first-best surplus across the two models. Recall that, in the standard order, $D(1) > 0$ is the condition under which a switching threshold $\theta$ exists in $[0,1]$. The interpretation: the Agent eventually adopts $H$ under some contract if and only if $H$ is the first-best efficient model. We write $m^{FB} = \arg\max_m U_m(1)$ for the first-best model, and
\[
S_m^{FB} \;\equiv\; U_m(1) \;=\; M_m\, h(\tau_m), \qquad h(z) \equiv 1 - z + z\ln z,
\]
for the first-best surplus achievable under model $m$.

\paragraph{Agency Cost} At the Principal's unconstrained optimum $\alpha_m^* = \sqrt{\tau_m}$, total surplus is $S_m(\sqrt{\tau_m}) = U_m(\sqrt{\tau_m}) + V_m(\sqrt{\tau_m})$, strictly below $S_m^{FB}$. The loss or cost of inducing Agent to select each model $m$ is given by:
\[
S_m^{FB} - S_m(\sqrt{\tau_m}) \;=\; M_m \sqrt{\tau_m}\, h(\sqrt{\tau_m}),
\]
and normalizing by first-best surplus gives a loss ratio that depends only on the activation threshold $\tau_m$,
\[
\ell_m \;\equiv\; \frac{S_m^{FB} - S_m(\sqrt{\tau_m})}{S_m^{FB}} \;=\; \frac{\sqrt{\tau_m}\, h(\sqrt{\tau_m})}{h(\tau_m)}.
\]

Further, re-arranging terms gives us:
\[
U_m(\sqrt{\tau_m}) \;=\; S_m^{FB} - S_m(\sqrt{\tau_m}).
\]
The Agent's utility at the Principal's optimal contract is exactly equal to the Agency cost, a consequence of the specific saturating-exponential functional form we assume; not a general property of moral hazard models.

\section{Burn-in Switching Threshold}
\label{app:burnin_same_family}
 
This appendix derives the qualitative behavior of the burn-in switching threshold $\theta_0$ (Eq.~\ref{eq:burnin_eq_gen}) for an illustrative special case. Motivated by calibration tables where $k$ and $b$ are roughly similar (not exactly equal) for models of the same family for a given task, we consider a special case with identical token saturation rates ($k_L = k_H = k$) and identical burn-in token requirements ($b_L = b_H = b$). Under this assumption, the participation thresholds (Eq.~\ref{eq:participation_threshold}) become constant scalar multiples of the corresponding activation thresholds, with a common scale factor $\rho = f^{-1}(kb) \ge 1$:
\[
\tau^P_L = \rho\,\tau_L, \qquad \tau^P_H = \rho\,\tau_H.
\]
 
\subsection{Flipped Order ($\tau_H < \tau_L$)}
 
When the high-capability model activates first, the baseline derivative condition remains intact, since the fixed burn-in costs vanish under differentiation:
\[
\mathcal{U}'_H(\alpha) = M_H\left(1 - \frac{\tau_H}{\alpha}\right) > M_L\left(1 - \frac{\tau_L}{\alpha}\right) = \mathcal{U}'_L(\alpha).
\]
Coupled with $\tau^P_H < \tau^P_L$ (which follows immediately from $\tau_H < \tau_L$ and the common scale factor $\rho$), model $H$ both breaks even earlier and climbs strictly faster than model $L$ at every point in the active domain. Consequently, model $H$ globally dominates, and no switching point exists exactly as in the base model without burn-in.
 
\begin{remark}
Without the same-family assumption, it is possible to have $\tau_H < \tau_L$ but $\tau^P_H > \tau^P_L$ (i.e., different $k_m, b_m$ break the common scale factor $\rho$). This would open a brief window in which $L$ is the only economically viable model, before $H$ eventually catches up and dominates.
\end{remark}
 
\subsection{Standard Order ($\tau_L < \tau_H$)}
 
When model $L$ activates first, the uniform scaling yields $\tau^P_L < \tau^P_H$. We reuse the baseline utility differential $D(\alpha)$ from Eq.~\eqref{eq:util_diff}, which is strictly convex under this ordering (since $M_H\tau_H > M_L\tau_L$ term-wise) with a unique zero $\theta$ on its increasing branch.
 
Rewriting Eq.~\eqref{eq:burnin_eq_gen} with common $b$:
\begin{equation}
    D(\theta_0) = (c_H - c_L)\,b.
    \label{eq:burnin_eq}
\end{equation}
Under the standing assumption $c_H > c_L$ (the more capable model carries a higher per token cost) and $b > 0$, the right-hand side is strictly positive, so
\[
    D(\theta_0) > D(\theta) = 0.
\]
Since $D$ is increasing on $(\theta, \infty)$ (the increasing branch of the convex function), $D(\theta_0) > D(\theta)$ directly implies
\[
    \theta_0 > \theta.
\]
The burn-in cost, being more expensive in absolute terms for the costlier model ($c_H b > c_L b$), strictly delays the switch. As in the base model, $\theta_0$ exists in $[0,1]$ only if $D(1) \geq (c_H - c_L)b$; otherwise $L$ dominates throughout the domain.
 
\section{Linear Contracts with Fixed Payments}
\label{app:fixed-payment}

We now consider a contract $\langle \alpha, \beta \rangle$, where $\alpha \in [0,1]$ remains the performance-based revenue share and $\beta \in \mathbb{R}$ is a flat transfer payment independent of output $q$. 

The updated normalized utilities for the Agent and Principal are:

\begin{align*}
U(\alpha, \beta, m, x) &= \alpha \cdot q_m(x) - c_m x + \beta, \\ V(\alpha, \beta, m, x) &= (1-\alpha) \cdot q_m(x) - \beta.
\end{align*}

\subsubsection{Optimal Effort and Model Choice}

The introduction of a fixed payment $\beta$ does not alter the marginal incentives for effort. Taking the first-order derivative of $U(\alpha, \beta, m, x)$ with respect to $x$ eliminates $\beta$, leaving the first-order condition identical to the pure revenue-sharing case. Consequently, the optimal effort $x^*_m$ and the activation threshold $\tau_m$ remain exactly as derived in Equations \eqref{eq:opt_effort} and \eqref{eq:opt_util}. 

Substituting optimal effort back into the Agent's utility gives:

\begin{align}
U_m(\alpha, \beta) &= U_m(\alpha) + \beta \nonumber \\ &= M_m\!\left[\alpha - \tau_m - \tau_m\ln\!\left(\frac{\alpha}{\tau_m}\right)\right] + \beta, \quad \alpha > \tau_m. \label{eq:opt_util_beta}
\end{align}

Similarly, because $\beta$ shifts the utility curves of all models $m \in \mathcal{M}$ by the same constant amount, it cancels out during the Agent's model selection phase. The condition $U_H(\alpha, \beta) > U_L(\alpha, \beta)$ is mathematically equivalent to $U_H(\alpha) > U_L(\alpha)$. Therefore, the switching threshold $\theta$ and the two scenarios (Standard vs. Flipped) remain entirely unchanged from the previous section.

\subsubsection{Principal's Optimization}

The presence of the fixed transfer fundamentally alters the Principal's optimization strategy. The Principal seeks to maximize $V_m(\alpha, \beta)$ subject to the Agent's Individual Rationality (IR) constraint, assuming an outside option utility of zero:
\[
    U_m(\alpha, \beta) \geq 0 \implies U_m(\alpha) + \beta \geq 0
\]

To maximize its own utility, the Principal will extract all surplus from the Agent by setting the fixed payment such that the IR constraint binds exactly:
\[
    \beta^* = - U_m(\alpha)
\]

If $U_m(\alpha) > 0$, this requires $\beta^* < 0$, functioning as a fee paid by the Agent to the Principal for the right to perform the task. 

Substituting $\beta^*$ into the Principal's objective function aligns the Principal's utility with the total social surplus of the system $S_m(\alpha)$:
\begin{align*}
    V_m(\alpha, \beta^*) &= (1-\alpha) \cdot q_m(x^*_m) - (-U_m(\alpha)) \\
    &= (1-\alpha) \cdot q_m(x^*_m) + \alpha \cdot q_m(x^*_m) - c_m x^*_m \\
    &= q_m(x^*_m) - c_m x^*_m \\
    &= S_m(\alpha)
\end{align*}

The Principal effectively designs the contract to maximize total surplus, which they then fully extract via $\beta^*$. Using $q_m(x^*_m) = M_m(1 - \tau_m/\alpha)$ and $x^*_m = \frac{1}{k_m}\ln\left(\frac{\alpha}{\tau_m}\right)$, the total surplus function is:
\[
    S_m(\alpha) = M_m\!\left(1 - \frac{\tau_m}{\alpha}\right) - \frac{c_m}{k_m}\ln\!\left(\frac{\alpha}{\tau_m}\right)
\]

Differentiating with respect to $\alpha$ and setting to zero:
\[
    \frac{dS_m}{d\alpha} = M_m \frac{\tau_m}{\alpha^2} - \frac{c_m}{k_m} \frac{1}{\alpha} = 0
\]

Recall from the activation threshold definition that $\frac{c_m}{k_m} = M_m \tau_m$.
Substituting this identity yields:
\[
    M_m \tau_m \left( \frac{1}{\alpha^2} - \frac{1}{\alpha} \right) = 0.
\]
Since $M_m > 0$ and $\tau_m > 0$, the only strictly positive solution is $\alpha=1$; also $dS_m/d\alpha > 0$ throughout $(0,1)$. This is the classic \emph{selling the firm} result: by setting $\alpha=1$, the Principal eliminates the misalignment between the Agent's private return and total surplus, since the Agent now keeps the full marginal return and bears the full
marginal cost of every token spent.

The Agent's model choice at $\alpha=1$ is governed by the switching threshold $\theta$ from previous section , independent of $\beta$. Since $S_m$ is increasing throughout $(0,1)$ for both models, $\alpha=1$ is surplus-maximizing and the sign of the utility differential $D(1)$ (see Eq.~\eqref{eq:util_diff}) identifies which model the Agent selects:
\[
    \alpha^*=1, \qquad m^* = \begin{cases} H & \text{if } D(1) > 0 \\ L & \text{if } D(1) \le 0. \end{cases}
\]
The Principal sets the fixed transfer to bind the Agent's IR constraint exactly, extracting
the realized surplus in full:
\[
    \beta^* = -U_{m^*}(1) = -M_{m^*}\!\left[1-\tau_{m^*} - \tau_{m^*}\ln\!\left(\frac{1}{\tau_{m^*}}\right)\right].
\]

\textit{Remark: } This implements the first-best outcome. At $\alpha=1$, the Agent's optimal effort (see Eq.~\eqref{eq:opt_effort}) coincides with the surplus-maximizing FOC a Principal with direct control over tokens would solve, so $x^*_m(1)$ is the efficient token budget. The fixed-payment contract therefore induces the efficient model and efficient effort simultaneously, with zero loss due to moral hazard (albeit trivially by selling the firm, which is not always realistic).

\subsubsection{Limited Liability Constraint ($\beta \geq 0$)}

If we assume limited liability, which is common in most real-world settings, it imposes the added constraint $\beta \geq 0$. The Principal solves:
\[
    \max_{\alpha, \beta} \; (1-\alpha) \cdot q_m(x^*_m(\alpha)) - \beta \quad \text{s.t.} \quad U_m(\alpha) + \beta \geq 0, \quad \beta \geq 0
\]

Because any $\beta > 0$ directly reduces the Principal's payoff without modifying the Agent's marginal incentives for effort, the limited liability constraint gives:
\[
    \beta^*_{LL} = 0
\]

Consequently, the problem collapses entirely back to the pure revenue-sharing model from previous section.

\section{Costly Verification}
\label{app:verification-cost}

We extend the base model to allow the Principal to incur a fixed cost $c_v \ge 0$ per round to observe (verify) the realized output quality $q$. The constant cost assumption is justified in cases such as math verification or unit tests for code outputs. In the base model this cost is implicitly zero; here we make it explicit. Note that we consider the case where the Principal verifies in every round. An interesting extension could be probabilistic verification, but we leave that for future work.

The verification cost $c_v$ is borne entirely by the Principal and, therefore, does not appear in the Agent's utility. Consequently, the Agent's optimal budget for model $m$ and the Agent's model choice through switching threshold $\theta$ remain unchanged. 

The Principal's utility from inducing model $m$ at share $\alpha$ becomes
\[
V_m^v(\alpha) \;=\; (1-\alpha)\,q_m(x_m^*(\alpha)) - c_v \;=\; V_m(\alpha) - c_v,
\]
i.e. the verified payoff is the unverified payoff $V_m(\alpha)$ shifted down by the constant fixed cost $c_v$.

For any $c_v \ge 0$ constant in $\alpha$, the FOC remains the same because the constant term gets differentiated out to zero. This implies the unconstrained optima $\sqrt{\tau_m}$ remains unchanged. We have shown that the model choice from Agent's side, determined by $\theta$ also does not change. Consequently the constrained optima $\alpha_L^\dagger$ and $\alpha_H^\dagger$, which are $\sqrt{\tau_m}$ clipped against the switching threshold $\theta$, are also unchanged.

So a constant verification cost changes \emph{none} of the base model's structural results: the Principal's optimal contract share is exactly as derived in the base model. Verification cost only ever shows up as a constant shift in the Principal's realized payoff. However, this matters for whether the Principal wants to contract at all. In the base model, the Principal's utility was always non-negative, but the fixed cost introduces the need to check participation constraint.

\subsection{Participation}

The Principal's payoff at the (unconstrained) optimum is
\[
V_m(\sqrt{\tau_m}) \;=\; M_m\big(1+\tau_m-\sqrt{\tau_m}-\tau_m/\sqrt{\tau_m}\big) \;=\; M_m\big(1 - \sqrt{\tau_m}\big)^2,
\]
The Principal participates \emph{i.e.}, offers a contract to induce model $m$ only if $V_m^v(\alpha_m^\dagger) \ge 0$,
\[
c_v \;\le\; V_m(\alpha_m^\dagger).
\]
At the unconstrained optimum this becomes the clean threshold $c_v \le M_m(1-\sqrt{\tau_m})^2$: a small verification cost relative to this never changes the Principal's optimal share, but a sufficiently large $c_v$ can make model $m$ non-viable for the Principal. 

\subsection{Alternative verification costs}

Treating $c_v$ as a fixed constant, while realistic in some scenarios, makes the verification problem fairly trivial as it gets differentiated out in all first-order conditions. Other interesting choices for verification cost are as follows.

\begin{itemize}
    \item \textbf{$c_v$ depending on $q$.} If verification cost scales with the quality being verified (e.g., low quality or buggy code will not even compile or error out quickly with the unit tests), $c_v = c_v(q)$ enters $V_m^v(\alpha)$ as a function of $\alpha$ through $q_m(x_m^*(\alpha))$, and the first-order condition picks up a $c_v'(q) \cdot q_m'(x_m^*)\cdot \frac{dx_m^*}{d\alpha}$ term and $\alpha_m^*$ would shift in general. 
    
    \item \textbf{$c_v$ depending on $m$.} Even holding $c_v$ constant in $\alpha$, allowing $c_v = c_{v,m}$ to differ by model (e.g. verifying a more elaborate $H$-model's output costs more than a short $L$-model output) does not change any of the structural results, but it does mean the participation thresholds in the previous subsection differ across models for a second reason (beyond $M_m, \tau_m$ already varying), which could be a source of an additional, verification-driven bias toward the cheaper-to-verify model.
    
    \item \textbf{Endogenous/probabilistic verification.} The Principal could choose to verify only with some probability $p < 1$, trading off expected verification cost $p\cdot c_v$ against reduced ability to enforce the contract (an Agent who anticipates low verification probability may deviate). This introduces strategic aspects to the interaction between verification and the Agent's incentives. 
\end{itemize}

\section{Experiment Details and Additional Results}

\paragraph{Datasets and Filtering.} For the MATH dataset~\cite{hendrycksmath2021}, which contains questions across difficulty levels 1 through 5, we filter exclusively for questions in levels 3 and 4 to maintain consistent task difficulty for calibrating production curves. We filter out questions containing graphic components by matching raw markup tags such as \texttt{[asy]}, resulting in $2,962$ questions for evaluation. The final JSON file is included in the code package. For MMLU-Pro~\cite{wang2024mmlu}, we evaluate across all $12,032$ available questions in the test partition.

\paragraph{Models, Serving, and Infrastructure.} All base models are loaded directly using their official Hugging Face repository identifiers, as listed in Table~\ref{tab:model_budgets}. All bandit runs for different pairings are executed on a single NVIDIA H100 GPU (80GB). Open-weights LLMs are served locally via vLLM using an OpenAI-compatible HTTP API server without additional model quantization. To ensure deterministic generation, all inference requests are made with zero sampling temperature ($\text{temperature} = 0.0$).

\paragraph{Answer Parsing and Evaluation.}
\begin{itemize}
    \item \textbf{MATH Dataset:} Evaluated using the \texttt{math\_verify} library to verify symbolic and numerical equivalence against the ground truth answer inside \texttt{\textbackslash boxed\{\}} outputs.
    \item \textbf{MMLU-Pro Dataset:} Using regex pattern matching to locate response strings matching \texttt{"answer is (X)"} or \texttt{"answer: (X)"}.
\end{itemize}
In both benchmarks, we first strip reasoning scratchpads (e.g., extracting text following \texttt{</think>} tags for DeepSeek models). Also, if a model generation is truncated due to budget constraints or fails answer extraction, it is assigned a accuracy score of $0$ ($0$ reward).

\paragraph{Action Space and Discretization.} Across both tasks, the action space for the bandit is discretized into $N = 21$ linearly spaced token bins spanning dataset- and model-specific minimum ($x_{\min}$) and maximum ($x_{\max}$) token limits.

\begin{table}[ht]
\centering
\small
\begin{tabular}{llcc}
\toprule
& & \multicolumn{2}{c}{($x_{\min}, x_{\max}$)} \\
\textbf{Model Family} & \textbf{Hugging Face Repository ID} & \textbf{MATH} & \textbf{MMLU-Pro} \\
\midrule
Llama 3.2 & \texttt{meta-llama/Llama-3.2-1B-Instruct} & $[100, 1000]$ & $[100, 1000]$ \\
          & \texttt{meta-llama/Llama-3.2-3B-Instruct} & $[100, 1000]$ & $[100, 1000]$ \\
\midrule
DeepSeek R1 & \texttt{deepseek-ai/DeepSeek-R1-Distill-Qwen-1.5B} & $[1000, 5000]$ & $[400, 5000]$ \\
          & \texttt{deepseek-ai/DeepSeek-R1-Distill-Qwen-7B}    & $[1000, 5000]$ & $[400, 5000]$ \\
\midrule
Gemma 4   & \texttt{google/gemma-4-E2B-it} & $[400, 3000]$ & $[400, 3000]$ \\
          & \texttt{google/gemma-4-E4B-it} & $[400, 3000]$ & $[400, 3000]$ \\
\bottomrule
\end{tabular}
\caption{Model identifiers and corresponding discretized token budget bounds ($x_{\min}, x_{\max}$)}
\label{tab:model_budgets}
\end{table}

\paragraph{Hyperparameters and Reproducibility.} All experiments use a single fixed random seed of $42$ ($\text{\texttt{seed}} = 42$). The decision agent runs for $2,000$ episodes on MATH and MMLU-Pro, using a batch size parameter of $k = 16$, i.e., in each round, the reward is the accuracy over a set of 16 questions. All code is attached as a zip. A more organized version will be made publicly available upon acceptance.

LinUCB is a contextual bandit algorithm~\cite{li_contextual-bandit_2010}: it assumes each arm's expected reward is a linear function of the current context, and maintains a running estimate of the weights (using ridge-regression) from observed rewards. Similar to UCB, at each round, it adds an exploration bonus which shrinks as more data accumulates for that arm. We use a smaller exploration weight at the model-choice level ($\gamma=1$) and a larger one at the token-budget level ($\gamma=2$): since the model-choice level's reward estimate depends on the token-budget policy already being close to its optimum, we use a larger exploration weight at the token-budget level, which has a larger action space (21 arms vs. 2), to ensure its reward estimates are reliable before the model-choice level's decision settles.
%converge faster, so that by the time the model-choice level settles, it is learning on reliable, near-optimal reward signals.
%Exploration control bounds are set to $\gamma_{\text{high}} = 1.0$ and $\gamma_{\text{low}} = 2.0$. 
%Non-linear least squares calibration fitting for parameters $(M_m, k_m, b_m)$ via \texttt{scipy.optimize.curve\_fit} uses standard unconstrained bounds initialized at $[0.5, 0.001, 100.0]$.

\begin{figure*}[htbp!] 
    \centering
    
    \begin{subfigure}[b]{0.31\textwidth}
        \centering
        \includegraphics[width=\textwidth]{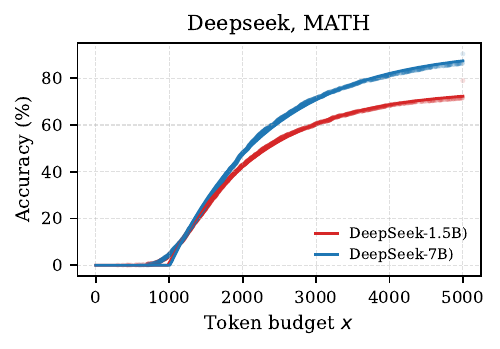}
    \end{subfigure}
    \hfill
    \begin{subfigure}[b]{0.31\textwidth}
        \centering
        \includegraphics[width=\textwidth]{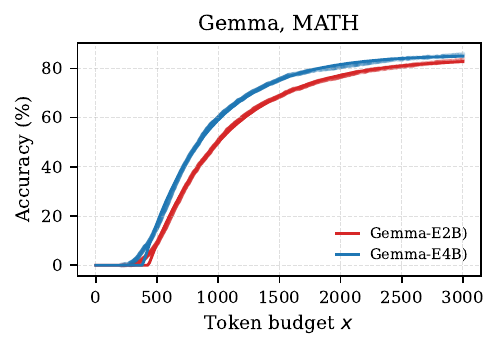}
    \end{subfigure}
    \hfill
    \begin{subfigure}[b]{0.31\textwidth}
        \centering
        \includegraphics[width=\textwidth]{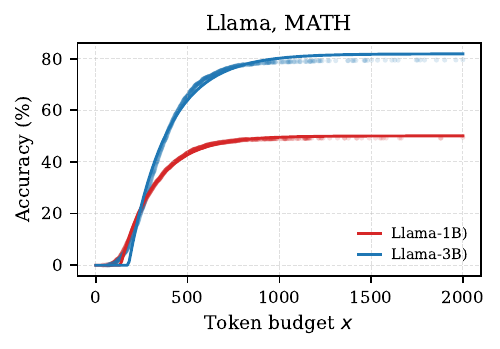}
    \end{subfigure}

     \begin{subfigure}[b]{0.31\textwidth}
        \centering
        \includegraphics[width=\textwidth]{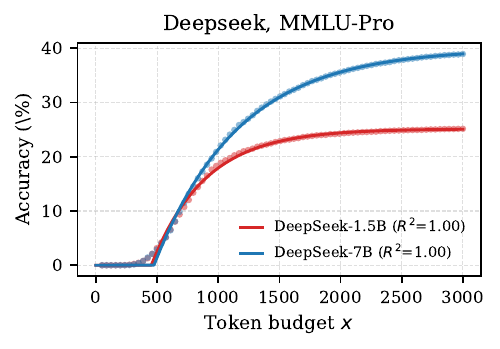}
    \end{subfigure}
    \hfill
    \begin{subfigure}[b]{0.31\textwidth}
        \centering
        \includegraphics[width=\textwidth]{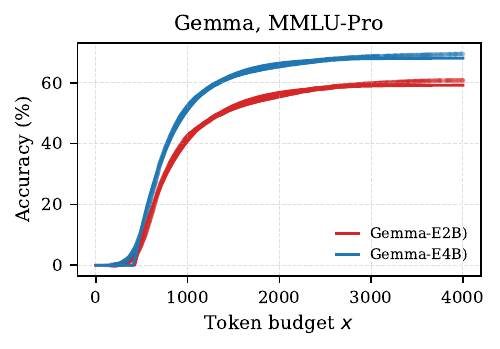}
    \end{subfigure}
    \hfill
    \begin{subfigure}[b]{0.31\textwidth}
        \centering
        \includegraphics[width=\textwidth]{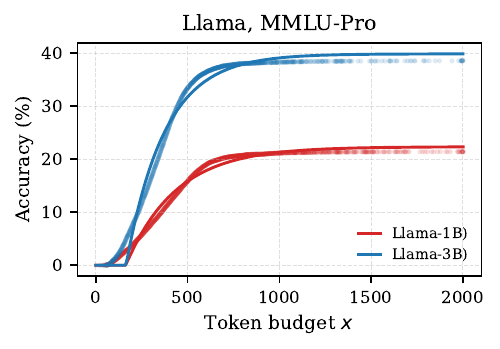}
    \end{subfigure}
    
    \caption{Accuracy vs budget for all 3 model families and 2 task domains, along with the fitted curves.}
    \label{fig:cal_matrix}
\end{figure*}

\begin{figure*}[htbp!] 
    \centering
    
    \begin{subfigure}[b]{0.48\textwidth}
        \centering
        \includegraphics[width=\textwidth]{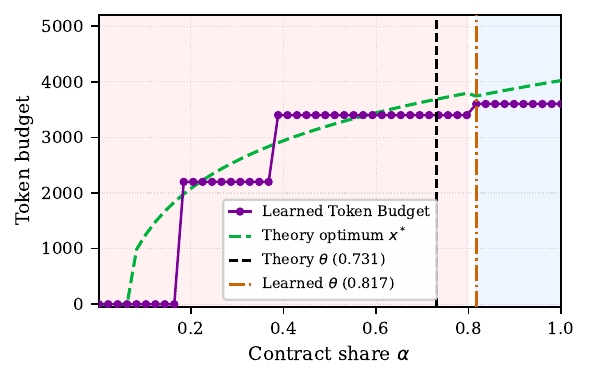}
        \caption{DeepSeek R1 1.5B \textit{vs.} 7B}
        \label{fig:math_ds_pure}
    \end{subfigure}
    \hfill
    \begin{subfigure}[b]{0.48\textwidth}
        \centering
        \includegraphics[width=\textwidth]{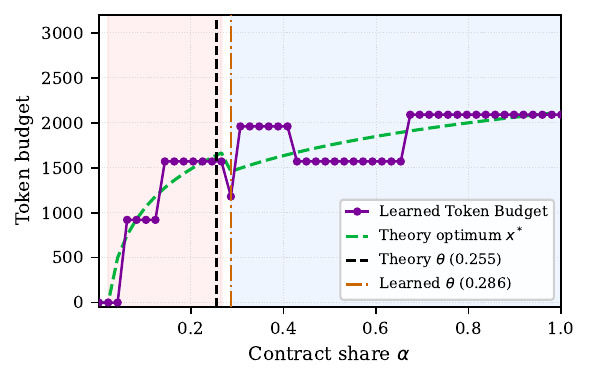}
        \caption{Gemma E2B \textit{vs.} E4B}
        \label{fig:math_gemma_pure}
    \end{subfigure}

    \begin{subfigure}[b]{0.48\textwidth}
        \centering
        \includegraphics[width=\textwidth]{Figures/math/learned_policy_llama_pure_math.pdf}
        \caption{Llama 1B \textit{vs.} 3B}
        \label{fig:math_llama_pure}
    \end{subfigure}
    \hfill
    \begin{subfigure}[b]{0.48\textwidth}
        \centering
        \includegraphics[width=\textwidth]{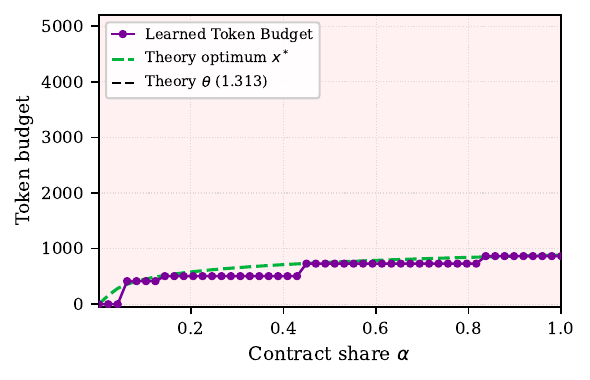}
        \caption{Llama 1B \textit{vs.} DeepSeek 1.5B}
        \label{fig:math_llama1b_vs_ds1_5b}
    \end{subfigure}

    \begin{subfigure}[b]{0.48\textwidth}
        \centering
        \includegraphics[width=\textwidth]{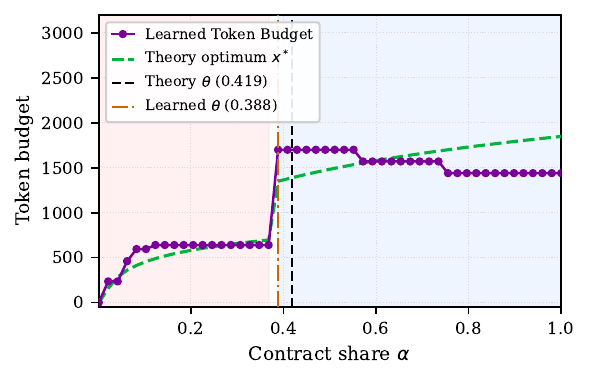}
        \caption{Llama 1B \textit{vs.} Gemma 4B}
        \label{fig:math_llama1b_vs_gemma4b}
    \end{subfigure}
    \hfill
    \begin{subfigure}[b]{0.48\textwidth}
        \centering
        \includegraphics[width=\textwidth]{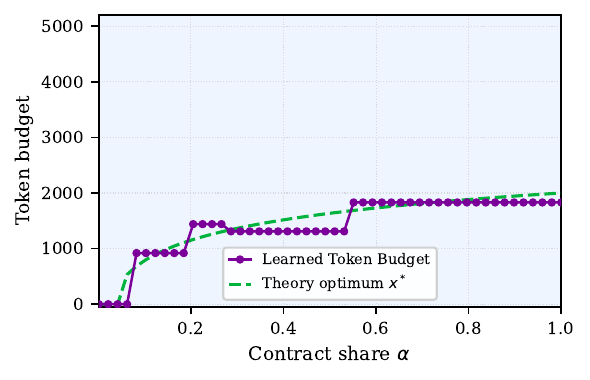}
        \caption{DeepSeek 1.5B \textit{vs.} Gemma 4B}
        \label{fig:math_ds1_5b_vs_gemma4b}
    \end{subfigure}

    \caption{Learned policies of various model pairings in the \texttt{MATH} domain. }
    \label{fig:math_policy_matrix}
\end{figure*}

\begin{figure}[p] 
    \centering
    
    \begin{subfigure}[b]{0.48\textwidth}
        \centering
        \includegraphics[width=\textwidth]{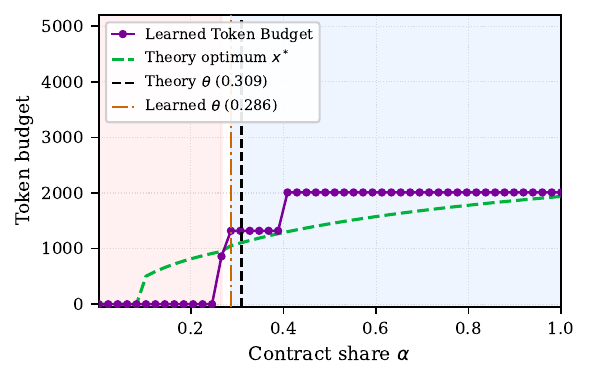}
        \caption{DeepSeek R1 1.5B \textit{vs.} 7B}
        \label{fig:ds_pure}
    \end{subfigure}
    \hfill
    \begin{subfigure}[b]{0.48\textwidth}
        \centering
        \includegraphics[width=\textwidth]{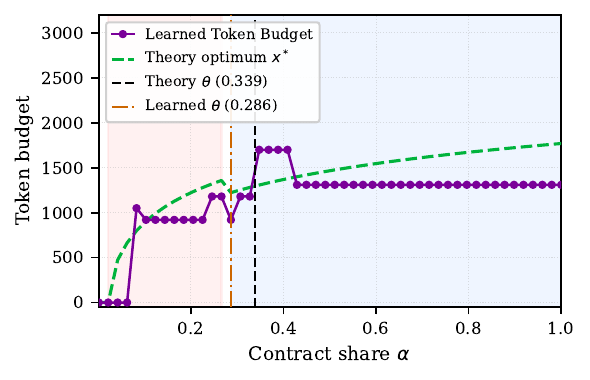}
        \caption{Gemma E2B \textit{vs.} E4B}
        \label{fig:gemma_pure}
    \end{subfigure}

    \begin{subfigure}[b]{0.48\textwidth}
        \centering
        \includegraphics[width=\textwidth]{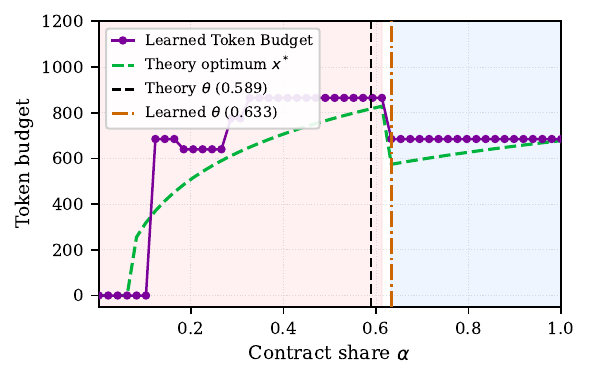}
        \caption{Llama 1B \textit{vs.} 3B}
        \label{fig:llama_pure}
    \end{subfigure}
    \hfill
    \begin{subfigure}[b]{0.48\textwidth}
        \centering
        \includegraphics[width=\textwidth]{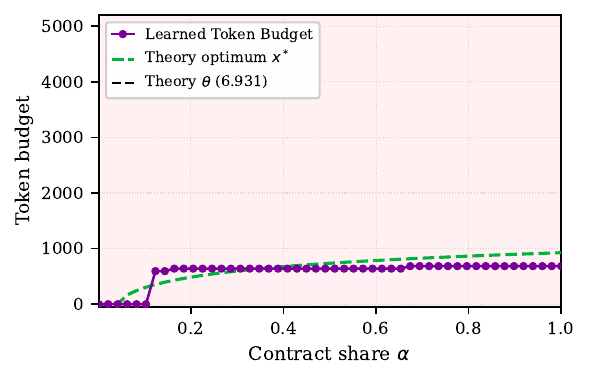}
        \caption{Llama 1B \textit{vs.} DeepSeek 1.5B}
        \label{fig:llama1b_vs_ds1_5b}
    \end{subfigure}

    \begin{subfigure}[b]{0.48\textwidth}
        \centering
        \includegraphics[width=\textwidth]{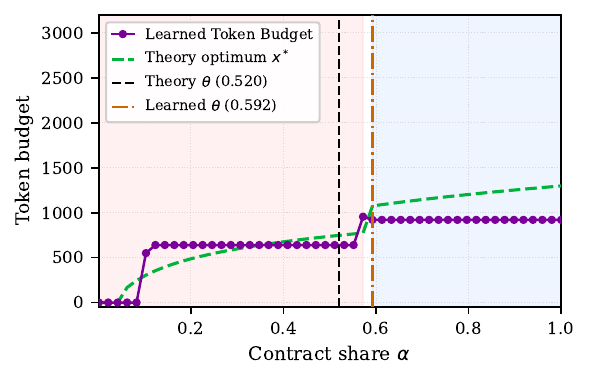}
        \caption{Llama 1B \textit{vs.} Gemma 4B}
        \label{fig:llama1b_vs_gemma4b}
    \end{subfigure}
    \hfill
    \begin{subfigure}[b]{0.48\textwidth}
        \centering
        \includegraphics[width=\textwidth]{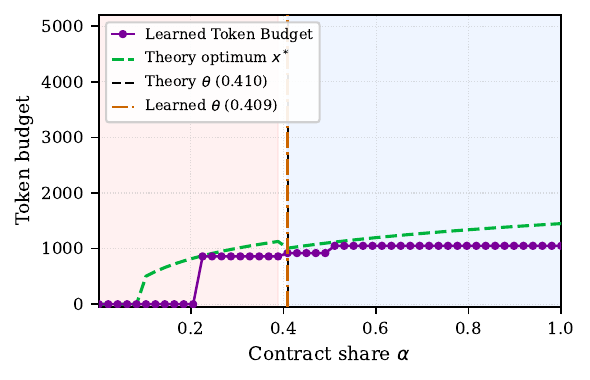}
        \caption{DeepSeek 1.5B \textit{vs.} Gemma 2B}
        \label{fig:ds1_5b_vs_gemma2b}
    \end{subfigure}

    \begin{subfigure}[b]{0.48\textwidth}
        \centering
        \includegraphics[width=\textwidth]{Figures/mmlu/learned_policy_ds1_5b_vs_gemma4b_mmlu.pdf}
        \caption{DeepSeek 1.5B \textit{vs.} Gemma 4B}
        \label{fig:ds1_5b_vs_gemma4b}
    \end{subfigure}
    \hfill
    \begin{subfigure}[b]{0.48\textwidth}
        \centering
        \includegraphics[width=\textwidth]{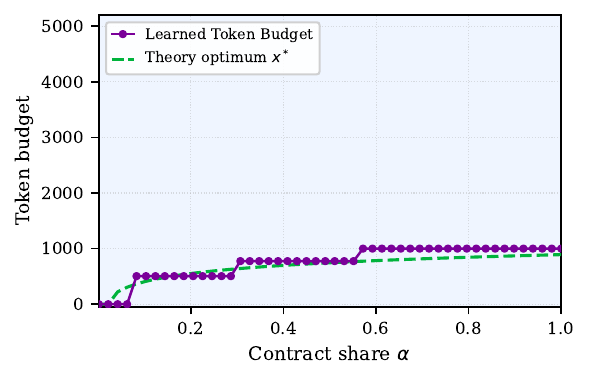}
        \caption{DeepSeek 1.5B \textit{vs.} Llama 3B}
        \label{fig:ds1_5b_vs_llama3b}
    \end{subfigure}

    \caption{Learned policies for the \texttt{MMLU Pro} domain.}
    \label{fig:mmlu_policy_matrix}
\end{figure}

\begin{figure*}[htbp!] 
    \centering
    
    \begin{subfigure}[b]{0.48\textwidth}
        \centering
        \includegraphics[width=\textwidth]{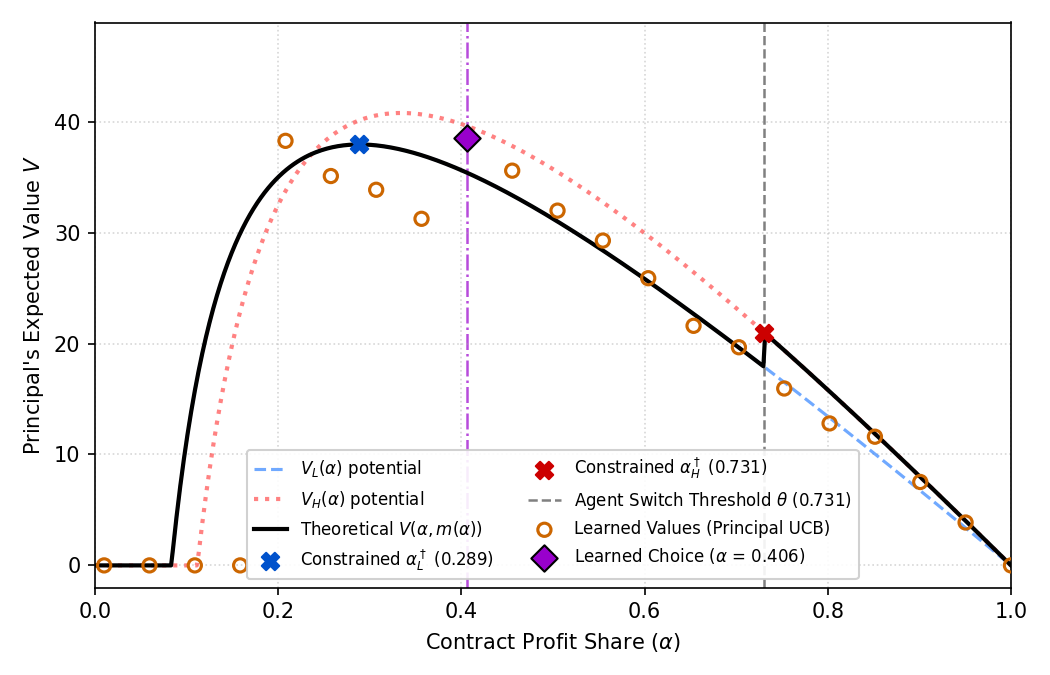}
        \caption{DeepSeek R1 1.5B \textit{vs.} 7B}
    \end{subfigure}
    \hfill
    \begin{subfigure}[b]{0.48\textwidth}
        \centering
        \includegraphics[width=\textwidth]{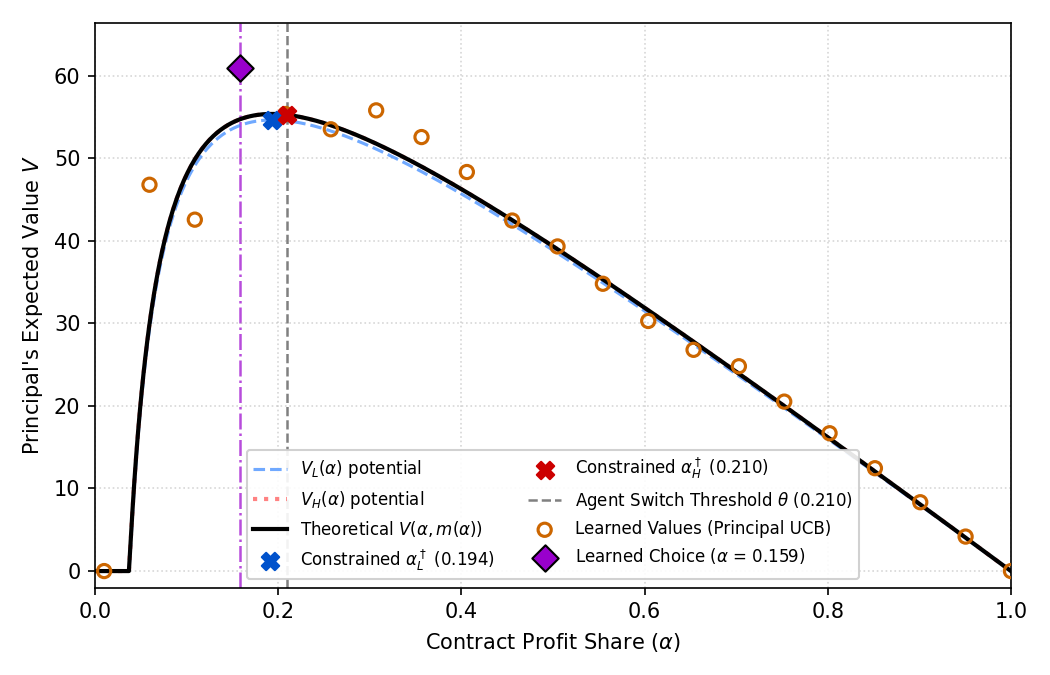}
        \caption{Gemma E2B \textit{vs.} E4B}
    \end{subfigure}

    \begin{subfigure}[b]{0.48\textwidth}
        \centering
        \includegraphics[width=\textwidth]{Figures/principal/llama.png}
        \caption{Llama 1B \textit{vs.} 3B}
    \end{subfigure}
    \hfill
    \begin{subfigure}[b]{0.48\textwidth}
        \centering
        \includegraphics[width=\textwidth]{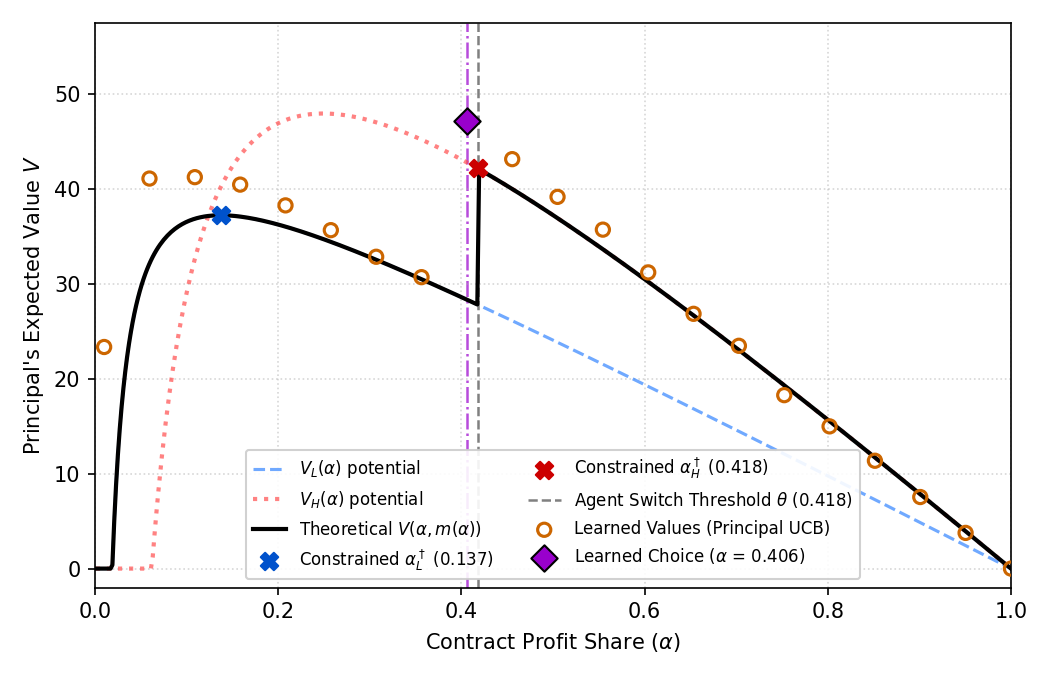}
        \caption{Llama 1B \textit{vs.} Gemma 4B}
    \end{subfigure}
    
    \caption{Principal's Learning for different model pairings on the MATH task}
    \label{fig:principal_matrix}
\end{figure*}

\begin{figure*}[htbp!] 
    \centering
    
    \begin{subfigure}[b]{0.48\textwidth}
        \centering
        \includegraphics[width=\textwidth]{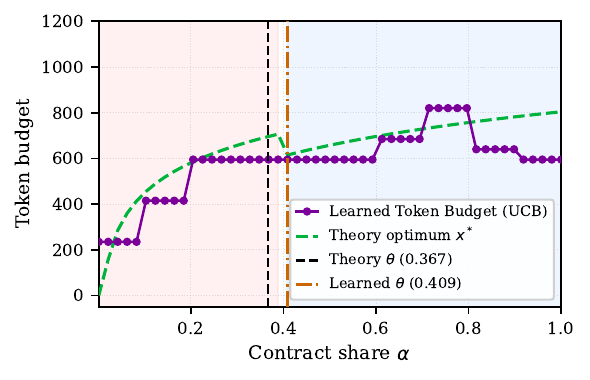}
        \caption{Llama 1B \textit{vs.} 3B}
    \end{subfigure}
    \hfill
    \begin{subfigure}[b]{0.48\textwidth}
        \centering
        \includegraphics[width=\textwidth]{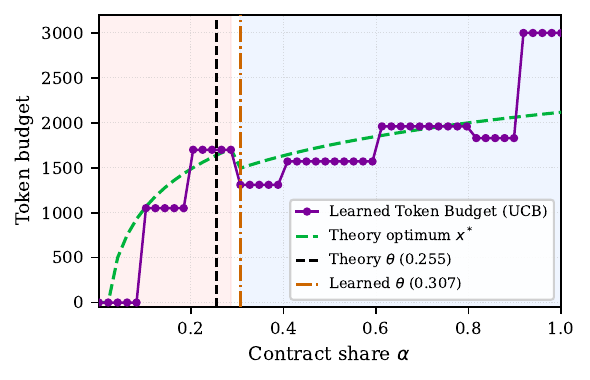}
        \caption{Gemma E2B \textit{vs.} E4B}
    \end{subfigure}
    \caption{Agent using simple UCB in MATH task for two different pairings. The context $\alpha$ is discretized into 10 bins, and we use a standard UCB instead of LinUCB with handcrafted features. The Agent considers each bin to be independent, and therefore takes a longer time to converge, and these were trained for $5,000$ episodes.}
    \label{fig:simple_ucb}
\end{figure*}
\clearpage

\section{Appendix F: LLM as Controller}
%\kate{You might need to move a lot of the text into the appendix, but it would be useful to just include a paragraph stating the results, to say "we thought about it, doesn't work. You need the infrastructure in place"}
Instead of the LinUCB learning Agent, we replace it with a LLM Controller, where we prompt a LLM with details about the general task domain, the 2 LLM models it has access to, their respective costs, and the profit objective it is supposed to maximize. Note that the prompt does not include any of our theory or calibrated parameters. We instead include a sliding window of the last $20$ decisions including details about the contract offered, the model chosen, token budget allocated, the resulting accuracy and reward. Finally, we append the contract offered in the current round, and prompt it to return a JSON object with model choice, token budget, and a justification. The exact prompt used is in Appendix G.

\begin{figure}[h]
    \centering
    \includegraphics[width=0.7\linewidth]{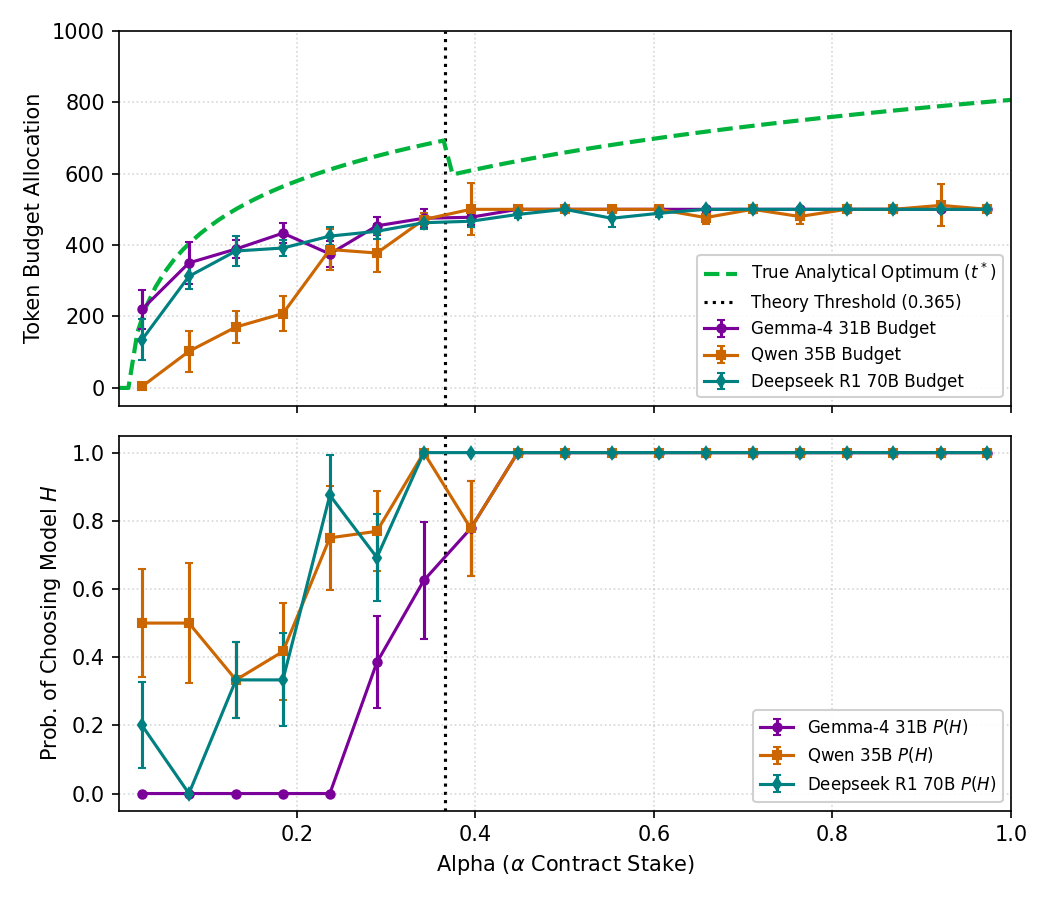}
    \caption{LLM as Controller}
    \label{fig:llm_controller}
\end{figure}

We experiment with a few different controller models; significantly more advanced and larger models compared to the model choices. We test it out for 220 rounds, with the first 20 rounds not being used in analysis as the sliding window of history is not yet full. We consider Gemma 4 31B, Qwen 3.6 35B, Deepseek R1 70B with zero temperature. We test this on the MATH domain and the Llama 1B \textit{vs.} 3B pairing, and the observed results are shown in Figure~\ref{fig:llm_controller}. Recall that the controller receives a contract stake $\alpha_t$ drawn uniformly at random from the domain $\alpha_t \in [0, 1]$. So, we discretize the contracts into 20 bins and average the model choices within each bin; the standard errors are also reported. 

All 3 controllers pick up the qualitative direction of model choice, with $P(H)$ rising with $\alpha$ and saturating near the theoretical threshold. The sharpness of this switch is not the same as the bandit controller and varies significantly within these models; the Qwen and Deepseek controllers pick model $H$ at higher rates than optimal below the threshold. On token budgets, all controllers spend less than optimal, with the token budget plateauing past the threshold. We take it as preliminary evidence that finding the Agent's two-dimensional best response is not trivial, and leave a systematic study of LLM controllers for future work. More importantly, this also does not account for the \emph{token cost of prompting the controller} itself each round, on top of the Agent's own token spend for the actual task; a cost the bandit-based approach does not incur.

\section{Appendix G: Prompts}
\label{app:prompts}

\begin{tcolorbox}[
    colback=gray!5, 
    colframe=gray!60, 
    title=\texttt{MATH} Prompt Template, 
    fonttitle=\bfseries,
    fontupper=\small\ttfamily,
    sharp corners,
    boxrule=0.5mm
]
\textbf{[System Instructions]} \\
Please reason step by step, and put your final answer within \textbackslash{}boxed\{\}.

\textbf{[User Payload]} \\
Question: [problem\_text]
\end{tcolorbox}

\begin{tcolorbox}[
    colback=gray!5, 
    colframe=gray!60, 
    title=\texttt{MMLU-Pro} Prompt Template, 
    fonttitle=\bfseries,
    fontupper=\small\ttfamily,
    sharp corners,
    boxrule=0.5mm
]
\textbf{[System Instructions]} \\
The following is a multiple-choice question (with answer) about [category]. Think step by step and then finish your answer with "The answer is (X)" where X is the correct letter choice.

\textbf{[User Payload]} \\
Question: [question\_text]

Options:
(A) [Option 1]
(B) [Option 2]
...
(N) [Option N]

Let's think step by step.
\end{tcolorbox}

\begin{tcolorbox}[
    colback=gray!5, 
    colframe=gray!60, 
    title=LLM Controller Prompt Template, 
    fonttitle=\bfseries,
    fontupper=\small\ttfamily,
    sharp corners,
    boxrule=0.5mm
]
\textbf{[System Instructions]} \\
You are an adaptive, economically optimal routing agent.

Context: In each round, you manage a batch of 16 mathematical problems sampled uniformly from a fixed distribution (MATH dataset, difficulty levels 3-4).

Objective: Maximize net profit per question = (alpha * accuracy\_pct) - (cost\_per\_token * token\_budget)

Configuration Space:

- Option A: '[Model Low Name]' (cost: [Cost Low])

- Option B: '[Model High Name]' (cost: [Cost High])

- Allowed Budget Range per question: [Token Min] to [Token Max]

--- START RECENT PERFORMANCE LOG ---

[History Window Log String, e.g., ``Round i: Alpha=... | Model=... | Budget=...``]

--- END RECENT PERFORMANCE LOG ---

Constraint: Respond ONLY with a valid JSON object matching this schema. Choose any integer value between [Token Min] and [Token Max] for token\_budget:

\{

\quad  "justification": "reasoning based on the objective, history and current alpha",
  
\quad  "chosen\_model": "model\_name",
  
\quad  "token\_budget": <integer value here>

\}

\textbf{[User Payload]} \\
Current Contract: alpha = <sampled alpha>
\end{tcolorbox}

\section{Appendix H: Extension to $N$ Models}

We consider $\mathcal{M} = \{1, \dots, N\}$ in place of $\{L, H\}$. The per-model quantities $\tau_m$, $x_m^*(\alpha)$, and $U_m(\alpha)$ are unchanged, since they follow from a single-model optimization and do not depend on the size of $\mathcal{M}$. What changes is the Agent's model choice, which is now
\[
m^*(\alpha) = \arg\max_{m \in \mathcal{M}} U_m(\alpha),
\]
i.e., the Agent's optimal model choice is based on the upper envelope of $N$ curves of the form~\eqref{eq:opt_util}, rather than a single crossing between two. We characterize this envelope in two steps: first, which models can be removed from consideration entirely; second, whether the remaining models are visited in capability order as $\alpha$ increases.

\subsection*{Capability Order}

Model $i$ is \emph{dominated} by model $j$ if $M_j \ge M_i$ and $\tau_j \le \tau_i$, with at least one inequality strict. We claim a dominated model is never the Agent's best response.

Suppose $\tau_j \le \tau_i$. For $\alpha > \tau_i$, both models are active and $\tau_j/\alpha \le \tau_i/\alpha$, so
\[
\left(1 - \frac{\tau_j}{\alpha}\right) \ge \left(1 - \frac{\tau_i}{\alpha}\right).
\]
Combined with $M_j \ge M_i$, this gives $U_j'(\alpha) \ge U_i'(\alpha)$ for all $\alpha > \tau_i$, by the same argument used for the Flipped Order case above. At $\alpha = \tau_i$, $U_i(\tau_i) = 0$ while $U_j(\tau_i) \ge 0$ (model $j$ is already active, having $\tau_j \le \tau_i$). Since $U_j$ starts weakly ahead of $U_i$ at $\alpha = \tau_i$ and climbs at least as fast for every $\alpha$ beyond it, $U_j(\alpha) \ge U_i(\alpha)$ for all $\alpha \ge \tau_i$; for $\alpha < \tau_i$, $U_i(\alpha) = 0 \le U_j(\alpha)$ trivially. So $U_j(\alpha) \ge U_i(\alpha)$ on all of $[0,1]$, and model $i$ never wins the envelope.

Removing dominated models, the remaining set $\{1, \dots, K\} \subseteq \mathcal{M}$ has no pair related this way. Sorting by capability, $M_{(1)} < \cdots < M_{(K)}$, we must also have
\[
\tau_{(1)} < \cdots < \tau_{(K)},
\]
since a violation would mean some pair is still dominated. We refer to this as \emph{capability order}: within the surviving set, more capable models are also more costly to activate.

Capability order alone does not imply the Agent moves through models $1, \dots, K$ one at a time as $\alpha$ increases. For $i < j$ define $D_{ij}(\alpha) = U_j(\alpha) - U_i(\alpha)$; as before, $D_{ij}$ is convex with at most one root $\theta_{ij}$. For three models $i<j<k$,
\[
D_{ik}(\alpha) = D_{ij}(\alpha) + D_{jk}(\alpha).
\]
If $\theta_{ij} \le \theta_{jk}$, this forces $\theta_{ik} \in [\theta_{ij}, \theta_{jk}]$ and model $j$ is optimal on that interval, as expected. But capability order does not guarantee $\theta_{ij} \le \theta_{jk}$; if instead $\theta_{ij} > \theta_{jk}$, model $j$ is never optimal on $[0,1]$ even though it is undominated and correctly placed in capability order, since the Agent prefers switching directly from $i$ to $k$. For example, consider
\[
(M_1,\tau_1) = (10, 0.1), \quad (M_2,\tau_2) = (11, 0.5), \quad (M_3,\tau_3) = (50, 0.6),
\]
which are in capability order and pairwise undominated. Direct computation gives $D_{12}(1) \approx -5.01 < 0$, so $\theta_{12} > 1$; and $D_{23}(1) \approx 2.99 > 0$, so $\theta_{23} \in (0,1)$. Since $\theta_{12} > \theta_{23}$, model 2 is never the Agent's choice for any $\alpha \in [0,1]$: it would only become optimal at a contract share past the allowed range. 

\subsection*{Staircase structure}

\textbf{Proposition.} Let models $1,\dots,K$ be undominated and in capability order. If the adjacent switching points satisfy
\[
\theta_{1,2} < \theta_{2,3} < \cdots < \theta_{K-1,K},
\]
then $m^*(\alpha) = i$ for $\alpha \in (\theta_{i-1,i}, \theta_{i,i+1})$, and the $K-1$ adjacent thresholds fully determine the envelope; the remaining pairwise comparisons are unnecessary.

The case $K=2$ is the base model. For the inductive step, suppose the ordering holds up to model $i-1$. Since $D_{i-1,i}$ is convex with a single root at $\theta_{i-1,i}$, once model $i$ overtakes model $i-1$ it remains ahead for all larger $\alpha$. Any earlier model $j<i-1$ is, by the induction hypothesis, already behind model $i-1$ once $\alpha$ exceeds $\theta_{i-2,i-1} < \theta_{i-1,i}$, and hence remains behind model $i$ as well by the same convexity argument applied to $D_{j,i}$. So checking neighbors suffices.

When the ordering condition fails, as in the example above, the envelope must be computed from all pairwise thresholds rather than adjacent ones alone; this is the standard problem of finding the upper envelope of pairwise-crossing curves.

\subsection*{Principal's Problem}

Given the ordering condition of the Proposition, the Principal's problem extends directly. On each interval $(\theta_{i-1,i},\theta_{i,i+1})$ where model $i$ is the Agent's choice, $V_i(\alpha)$ remains single-peaked at $\sqrt{\tau_i}$ exactly as in ~\eqref{eq:alpha_dagger}-\eqref{eq:principals_opt}, so
\[
\alpha_i^\dagger = \mathrm{clip}\!\left(\sqrt{\tau_i},\ \theta_{i-1,i},\ \theta_{i,i+1}\right), \qquad
\alpha^* = \arg\max_i V_i(\alpha_i^\dagger).
\]
No new derivation is required beyond the base model once the interval structure is known.

\end{document}